\documentclass[manuscript,screen,nonacm]{acmart}

\usepackage{longtable}
\usepackage{capt-of}
\usepackage{balance,bm}
\usepackage{color, colortbl, xcolor}
\definecolor{LightGray}{gray}{0.97}
\usepackage{pgfplots}
\usepackage{xcolor}
\usepackage{url}
\usepackage{makecell}
\usepackage{tabularx}
\usepackage{longtable}

\usepackage{subcaption}
\usepackage{booktabs} 
\usepackage{graphicx}
\usepackage{textcomp}
\usepackage{color,soul}
\usepackage{bm}
\usepackage{multirow}
\usepackage{wrapfig}
\usepackage{balance}
\usepackage{graphicx}  
\usepackage{enumitem}

\usepackage{colortbl}
\usepackage{arydshln}
\usepackage [english]{babel}
\usepackage [autostyle, english = american]{csquotes}
\definecolor{linkColor}{RGB}{6,125,233}
\definecolor{green}{rgb}{0.0, 0.65, 0.31}
\definecolor{bleudefrance}{rgb}{0.19, 0.55, 0.91}
\definecolor{ceruleanblue}{rgb}{0.16, 0.32, 0.75}
\definecolor{grey}{HTML}{969696}
\definecolor{violet}{HTML}{756bb1}
\definecolor{dgrey}{HTML}{01665e}
\definecolor{lgrey}{HTML}{5ab4ac}
\definecolor{dgreen}{HTML}{005a32}
\definecolor{purple}{HTML}{ae017e}

\definecolor{editCol}{HTML}{000000}
\definecolor{maskCol}{HTML}{c51b7d}
\definecolor{lrColor}{HTML}{8856a7}
\definecolor{trColor}{HTML}{d01c8b}
\definecolor{ctColor}{HTML}{4dac26}
\definecolor{brickred}{HTML}{f03b20}
\definecolor{improveCol}{HTML}{253494}
\definecolor{worsenCol}{HTML}{d7191c}
\definecolor{DarkBlue}{HTML}{00008B}
\definecolor{mscolor}{HTML}{01665e}
\definecolor{nmscolor}{HTML}{bf812d}
\definecolor{lgreen}{HTML}{ccece6}
\definecolor{dolive}{HTML}{308014}

\definecolor{lrColor}{HTML}{8856a7}
\definecolor{trColor}{HTML}{d01c8b}
\definecolor{ctColor}{HTML}{4dac26}
\definecolor{brickred}{HTML}{f03b20}
\definecolor{improveCol}{HTML}{253494}
\definecolor{worsenCol}{HTML}{d7191c}
\definecolor{lgreen}{HTML}{e0f3db}
\definecolor{dpink}{HTML}{CD1076}
\definecolor{pink}{HTML}{FED2D2}
\definecolor{soothinggreen}{HTML}{4dac26}
\definecolor{darkred}{HTML}{8B0000}

\definecolor{dblue}{HTML}{104E8B}
\definecolor{violet}{HTML}{8A2BE2}
\definecolor{mscolor}{HTML}{01665e}
\definecolor{nmscolor}{HTML}{d8b365}
\definecolor{deepgrey}{HTML}{525252}
\definecolor{dslate}{HTML}{2F4F4F}
\definecolor{dolive}{HTML}{556B2F}
\definecolor{teal}{HTML}{388E8E}
\definecolor{mscolor}{HTML}{01665e}
\definecolor{nmscolor}{HTML}{d8b365}

\definecolor{aicolor}{HTML}{018571}
\definecolor{occolor}{HTML}{ff7799}

\definecolor{srcolor}{HTML}{e34a33}
\definecolor{smcolor}{HTML}{253494}
\definecolor{srsmcolor}{HTML}{7fcdbb}
\definecolor{bothcolor}{HTML}{fe9929}
\definecolor{onecolor}{HTML}{018571}
\definecolor{marroon}{HTML}{881c1c}

\usepackage{mathtools}

\usepackage{amsmath}
\usepackage{array}
\usepackage{xcolor}
\usepackage{arydshln}
\usepackage{siunitx}
\colorlet{tablerowcolor4}{gray!50} 

\newcommand*{\textlabel}[2]{%
  \edef\@currentlabel{#1}
  \phantomsection
  #1\label{#2}
}

\colorlet{tableheadcolor}{gray!25} 
\colorlet{tablerowcolor}{gray!15} 
\colorlet{tablerowcolor2}{gray!45} 
\colorlet{tablerowcolor3}{gray!25} 

\newcommand{\rowcollight}{\rowcolor{LightGray}} %
\usepackage{capt-of}
\usepackage{balance,bm}
\usepackage{color, colortbl, xcolor}

\usepackage{booktabs}  
\usepackage{url}
\usepackage{hyperref}
\hypersetup{
    colorlinks=true,
}

\usepackage{colortbl}
\usepackage{arydshln}
\usepackage [english]{babel}
\usepackage [autostyle, english = american]{csquotes}
\definecolor{linkColor}{RGB}{6,125,233}
\definecolor{green}{rgb}{0.0, 0.65, 0.31}
\definecolor{bleudefrance}{rgb}{0.19, 0.55, 0.91}
\definecolor{ceruleanblue}{rgb}{0.16, 0.32, 0.75}
\definecolor{grey}{HTML}{969696}
\definecolor{violet}{HTML}{756bb1}
\definecolor{dgrey}{HTML}{01665e}
\definecolor{lgrey}{HTML}{5ab4ac}
\definecolor{dgreen}{HTML}{005a32}
\definecolor{purple}{HTML}{ae017e}

\definecolor{editCol}{HTML}{000000}
\definecolor{maskCol}{HTML}{c51b7d}
\definecolor{lrColor}{HTML}{8856a7}
\definecolor{trColor}{HTML}{d01c8b}
\definecolor{ctColor}{HTML}{4dac26}
\definecolor{brickred}{HTML}{f03b20}
\definecolor{improveCol}{HTML}{253494}
\definecolor{worsenCol}{HTML}{d7191c}
\definecolor{DarkBlue}{HTML}{00008B}
\definecolor{mscolor}{HTML}{01665e}
\definecolor{nmscolor}{HTML}{bf812d}
\definecolor{lgreen}{HTML}{ccece6}
\definecolor{dolive}{HTML}{308014}

\definecolor{maskCol}{HTML}{c51b7d}
\definecolor{lrColor}{HTML}{8856a7}
\definecolor{trColor}{HTML}{d01c8b}
\definecolor{ctColor}{HTML}{4dac26}
\definecolor{brickred}{HTML}{f03b20}
\definecolor{improveCol}{HTML}{253494}
\definecolor{worsenCol}{HTML}{d7191c}
\definecolor{lgreen}{HTML}{e0f3db}
\definecolor{dpink}{HTML}{CD1076}
\definecolor{pink}{HTML}{FED2D2}
\definecolor{soothinggreen}{HTML}{4dac26}
\definecolor{darkred}{HTML}{8B0000}

\definecolor{dblue}{HTML}{104E8B}
\definecolor{violet}{HTML}{8A2BE2}
\definecolor{mscolor}{HTML}{01665e}
\definecolor{nmscolor}{HTML}{d8b365}
\definecolor{deepgrey}{HTML}{525252}
\definecolor{dslate}{HTML}{2F4F4F}
\definecolor{dolive}{HTML}{556B2F}
\definecolor{teal}{HTML}{388E8E}
\definecolor{mscolor}{HTML}{01665e}
\definecolor{nmscolor}{HTML}{d8b365}

\definecolor{aicolor}{HTML}{018571}
\definecolor{occolor}{HTML}{ff7799}

\definecolor{srcolor}{HTML}{e34a33}
\definecolor{smcolor}{HTML}{253494}
\definecolor{srsmcolor}{HTML}{7fcdbb}
\definecolor{bothcolor}{HTML}{fe9929}
\definecolor{onecolor}{HTML}{018571}
\definecolor{marroon}{HTML}{881c1c}

\usepackage{mathtools}

\usepackage{amsmath}
\usepackage{array}
\usepackage{xcolor}
\usepackage{arydshln}
\usepackage{siunitx}
\colorlet{tablerowcolor4}{gray!50} 

\usepackage{tcolorbox}

\colorlet{tableheadcolor}{gray!25} 
\colorlet{tablerowcolor}{gray!15} 
\colorlet{tablerowcolor2}{gray!45} 
\colorlet{tablerowcolor3}{gray!25} 

\newif{\ifhidecomments}
  \hidecommentsfalse 
\ifhidecomments
    \newcommand{\keran}[1]{}
    \newcommand{\melissa}[1]{}
    \newcommand{\dongwhi}[1]{}
    \newcommand{\koustuv}[1]{}
    \newcommand{\ravi}[1]{}
\else
    \newcommand{\keran}[1]{\textbf{\small\sffamily{\textcolor{DarkBlue}{[#1 -- Keran]}}}}
    \newcommand{\melissa}[1]{\textbf{\small\sffamily{\textcolor{dolive}{[#1 -- Melissa]}}}}
    \newcommand{\ravi}[1]{\textbf{\small\sffamily{\textcolor{marroon}{[#1 -- Ravi]}}}}
    \newcommand{\dongwhi}[1]{\textbf{\small\sffamily{\textcolor{dpink}{[#1 -- Dong Whi]}}}}
    \newcommand{\koustuv}[1]{\textbf{\small\sffamily{\textcolor{violet}{[#1 -- Koustuv]}}}}
  \fi
\usepackage{xcolor}

\newcommand{\edit}[1]{#1}

\newcommand{\cmr}{\textsc{CareMirror}}

\renewcommand{\textrightarrow}{$\rightarrow$}

\colorlet{tableheadcolor}{gray!25} 
\colorlet{tablerowcolor}{gray!5} 

\definecolor{neutralCol}{HTML}{dd1c77}
\definecolor{neutralGreen}{HTML}{31a354}
\definecolor{NewBlue}{HTML}{1879ba}
\definecolor{bleudefrance}{rgb}{0.19, 0.55, 0.91}  
\definecolor{AfTrColor}{HTML}{0868ac}  
\definecolor{BfTrColor}{HTML}{a8ddb5}  

\definecolor{AfCtColor}{HTML}{b10026}  
\definecolor{BfCtColor}{HTML}{fd8d3c}

\graphicspath{ {figures/} }

\newcommand{\para}[1]{\vspace{0.3em}\noindent\textbf{#1}~}

\AtBeginDocument{%
  \providecommand\BibTeX{{%
    \normalfont B\kern-0.5em{\scshape i\kern-0.25em b}\kern-0.8em\TeX}}}

\begin{document}


\title[\cmr{}: Bringing Caregiver Wellbeing into the Dementia Care Ecosystem]{\cmr{}: Bringing Caregiver Wellbeing into the Dementia Care Ecosystem}

\author{Jiayue Melissa Shi}
\orcid{0009-0007-0624-2421}
\affiliation{%
  \institution{University of Illinois Urbana-Champaign}
 \city{Urbana}
 \state{IL}
 \country{USA}}
 \email{mshi24@illinois.edu}

\author{Ethan Nguyen}
\orcid{0009-0009-6638-7715}
\affiliation{%
 \institution{University of Illinois Urbana-Champaign}
 \city{Urbana}
 \state{IL}
 \country{USA}}
 \email{enguy20@illinois.edu}

\author{Drishti Goel}
\orcid{0009-0000-6713-9240}
\affiliation{%
 \institution{University of Illinois Urbana-Champaign}
 \city{Urbana}
 \state{IL}
 \country{USA}}
 \email{drishti4@illinois.edu}

 \author{Upasana Natarajan}
\orcid{0009-0004-2097-6833}
\affiliation{%
 \institution{University of Illinois Urbana-Champaign}
 \city{Urbana}
 \state{IL}
 \country{USA}}
 \email{un4@illinois.edu}

 \author{Shashwat Srivatsa}
\orcid{0009-0001-3467-1721}
\affiliation{%
 \institution{University of Illinois Urbana-Champaign}
 \city{Urbana}
 \state{IL}
 \country{USA}}
 \email{ss262@illinois.edu}

\author{Daniel S. Brown}
\orcid{0000-0001-9919-4869}
\affiliation{%
 \institution{OSF HealthCare}
 \city{Peoria}
 \state{IL}
 \country{USA}}
 \email{daniel.s.brown@osfhealthcare.org}

\author{Violeta J. Rodriguez}
\orcid{0000-0001-8543-2061}
\affiliation{%
 \institution{University of Illinois Urbana-Champaign}
 \city{Champaign}
 \state{IL}
 \country{USA}}
 \email{vjrodrig@illinois.edu}

\author{Dong Whi Yoo}
\orcid{0000-0003-2738-1096}
\affiliation{%
 \institution{Indiana University Indianapolis}
 \city{Indianapolis}
 \state{IN}
 \country{USA}}
 \email{dy22@iu.edu}

\author{Ravi Karkar}
\orcid{0000-0003-1467-4439}
\affiliation{%
 \institution{University of Massachusetts Amherst}
 \city{Amherst}
 \state{MA}
 \country{USA}}
 \email{rkarkar@umass.edu}

\author{Koustuv Saha}
\orcid{0000-0002-8872-2934}
\affiliation{%
 \institution{University of Illinois Urbana-Champaign}
 \city{Urbana}
 \state{IL}
 \country{USA}}
 \email{ksaha2@illinois.edu}

\renewcommand{\shortauthors}{}



\begin{abstract}

Family caregivers of people living with dementia shoulder emotional and practical responsibilities, yet their own wellbeing often remains peripheral to dementia care. We present \cmr{}, an envisioned caregiver wellbeing ecosystem with interconnected caregiver- and clinician-facing interfaces for longitudinal reflection, personalized support, and caregiver-controlled sharing with clinical care teams. We conducted semi-structured interviews with 14 caregivers, using \cmr{} as a design probe to examine how they perceived this ecosystem and what expectations, concerns, and boundaries emerged around clinical connection. Caregivers valued attention to their wellbeing, longitudinal awareness, context-sensitive support, and clinical visibility when it could lead to meaningful follow-up. However, repeated reflection could become burdensome or emotionally difficult. Automatic clinical sharing could inhibit candid disclosure, and participants wanted control over what information entered clinical care. They also expected AI to support reflection and communication without replacing caregiver voice or clinician judgment. These findings affirm the value of bringing caregiver wellbeing into clinical care, while revealing that this connection changes how caregivers engage with such a system. We contribute design considerations for technologies that proactively bring caregiver wellbeing into clinical care settings without compromising caregiver agency and privacy.

\end{abstract}



\maketitle



\section{Introduction}
Dementia progressively affects memory, reasoning, and the ability to manage everyday life, often over the course of many years~\cite{kim2021scoping}. In the U.S., approximately one in nine adults aged 65 and older lives with dementia~\cite{rajan2021population}. Supporting this population relies heavily on family caregivers: in 2023, 11.5 million caregivers of people living with dementia provided an estimated 18.4 billion hours of unpaid care, averaging nearly 31 hours per week~\cite{rabarison2018economic}.
Dementia therefore affects not only the people diagnosed, but also the millions of family members who provide much of their day-to-day care~\cite{alzheimer2005alzheimer}. 
Family caregivers assume substantial responsibilities, including responding to behavioral and cognitive changes, coordinating healthcare, assisting with daily activities, and making complex care decisions~\cite{garcia2011anxiety,grunfeld1997caring}.
These continued responsibilities can substantially affect caregivers' own wellbeing, contributing to emotional distress, anxiety, depression, social isolation, and burnout~\cite{brodaty2009family,garcia2011anxiety,manzini2020emotional,shi2025balancing}.
At the same time, caregivers may have limited time or capacity to attend to their own needs while managing the demands of caregiving~\cite{queluz2020understanding,steenfeldt2021becoming,shi2025balancing}.
This creates an important tension in dementia care: caregivers are integral to supporting people living with dementia, yet their own wellbeing can remain peripheral to the systems of care in which they participate.
Supporting dementia caregivers therefore requires recognizing them not only as \textit{care partners}, but also as individuals with wellbeing needs of their own.

Technology has increasingly been used to support dementia caregiving through information provision, monitoring, skills training, care coordination, and other forms of caregiver assistance~\cite{guan2021taking,rettinger2020mixed,shin2022effects}. 
Other caregiver-facing technologies focus more directly on caregivers' own needs, providing psychoeducation, peer support, self-management resources, and mental health support~\cite{cheng2024effectiveness,shin2022effects,kim2024opportunities}. 
However, these efforts often address two sides of caregiving separately. 
Technologies situated within dementia care commonly help caregivers support the person living with dementia, whereas technologies addressing self-wellbeing are often offered as standalone resources or interventions~\cite{smriti2024emotion,shi2025balancing,cheng2024effectiveness}. 
As a result, caregivers' everyday wellbeing experiences may remain disconnected from the dementia care ecosystem~\cite{shi2025balancing,yuan2025supporting}. 

This separation is especially consequential because caregiver wellbeing is not static. 
Caregivers' emotional and practical needs can change as dementia progresses, caregiving responsibilities shift, and new challenges emerge~\cite{rettinger2020mixed,wawrziczny2016needing,shi2025balancing}. 
Yet, support for caregivers is often provided through discrete resources or interventions that caregivers must seek out when they recognize a need~\cite{cheng2024effectiveness,shin2022effects,shi2025balancing}. 
This places much of the work of noticing changes, interpreting their significance, and finding appropriate support on caregivers themselves, even when their time and attention are already constrained by caregiving responsibilities~\cite{queluz2020understanding,steenfeldt2021becoming}. 
At the same time, clinicians involved in dementia care may have limited insight into how caregivers are doing between clinical encounters, and caregiver wellbeing may remain outside routine clinical conversations~\cite{bhat2023we,shi2025balancing}. 
Together, these challenges point to an opportunity for technologies that do more than provide isolated resources: they could help caregivers follow their wellbeing over time, connect changing needs with relevant support, and create pathways for those needs to become visible within clinical care when appropriate.

Creating such a connection, however, raises design questions that cannot be resolved by technical integration alone. 
Caregiver wellbeing information originates in a personal context, while clinical care introduces different expectations around access, interpretation, and response. 
Therefore, a system that connects these settings must consider how caregivers want to engage with ongoing wellbeing support, what role they want clinicians to play, and how information should move between personal and clinical contexts. 
These questions are particularly important to examine from caregivers' perspectives before such technologies are integrated into real-world care.

To make these possibilities more concrete, we designed \cmr{}, a caregiver wellbeing system designed in collaboration with clinician coauthors. 
\cmr{} brings together interconnected caregiver- and clinician-facing interfaces that support wellbeing assessment and reflection over time, personalized resources and AI-assisted support, and potential communication of caregiver wellbeing information to clinical care teams. 
Rather than proposing \cmr{} as a finalized model for caregiver support, we use it as a design probe~\cite{hutchinson2003technology} through which caregivers can experience and critique one possible way of connecting everyday wellbeing support with clinical care.
Accordingly, we investigate the following research questions (RQs):

\para{RQ1:} How do dementia caregivers perceive and critique \cmr{}'s design for supporting their wellbeing?

\para{RQ2:} What expectations and concerns do dementia caregivers have about their wellbeing being integrated into the dementia care ecosystem?

We conducted a semi-structured interview study with 14 caregivers. 
Participants first discussed their caregiving experiences and wellbeing needs, and then interacted with \cmr{} through a guided exploration of its wellbeing assessments, mood check-ins, longitudinal features, AI chatbot, and caregiver resources. 
We used \cmr{} to elicit how they understood the proposed features, what they found useful or limiting, and how they imagined these features fitting into their caregiving experiences. 
We then broadened the discussion beyond the current interface, asking participants to consider how caregiver wellbeing support might connect with clinical care, including what information they would want clinicians to receive, how sharing should occur, and what roles technology and clinicians should play. 
Participants additionally completed the System Usability Scale (SUS)~\cite{brooke1996sus} and Intervention Appropriateness Measure (IAM)~\cite{weiner2017psychometric}. 
\cmr{} received a mean SUS score of 81.61 (median=80.0) and a mean IAM score of 17.43 (median=18.0), indicating favorable perceptions of usability and appropriateness within this prototype-based study.

Qualitatively, participants valued creating an explicit space for caregiver wellbeing and saw value in following their wellbeing over time, particularly when support reflected their current caregiving circumstances. 
At the same time, participants identified limits to these forms of support: repeated wellbeing activities could become another demand, and revisiting difficult experiences was not always beneficial. 
Participants also saw value in making caregiver wellbeing visible to clinicians when doing so could lead to meaningful follow-up, while emphasizing that clinical access should not be an automatic consequence of using the caregiver-facing system. 
They wanted more control over what information moved into clinical care and expressed concerns about AI interpreting or communicating their experiences without sufficient human oversight. 
These findings move beyond evaluating individual interface features to show how caregiver wellbeing technologies change when personal reflection and support become connected with a broader care ecosystem. 
In particular, the findings suggest that such systems must negotiate the relationship between caregiver reflection, clinical visibility, caregiver agency, and professional involvement. 
These themes reflect the central tensions developed across the current Findings section. 

Our work contributes: (1) the design of \cmr{}, an interconnected caregiver- and clinician-facing prototype that shows one possible wellbeing ecosystem for dementia caregivers; (2) empirical insights into how dementia caregivers perceive the design of longitudinal and personalized wellbeing support and the opportunities and tensions that emerge when such support is connected with clinical care; and (3) design considerations for caregiver wellbeing technologies that reduce added burden, support meaningful longitudinal reflection, preserve caregiver control over clinical sharing, and position technology as supporting rather than replacing caregiver expression and clinician judgment.

\section{Related Work}

\subsection{Dementia: Condition and Caregiving}

Family caregivers assume substantial responsibilities throughout dementia care, including assisting with activities of daily living, responding to behavioral and cognitive changes, coordinating healthcare, managing safety concerns, and making complex decisions about ongoing and future care~\cite{brodaty2009family,harper2022,alzheimer2005alzheimer,rettinger2020mixed,queluz2020understanding,lee2022unmet,wang2026taxonomy}. 
These responsibilities can substantially affect caregivers' own wellbeing, contributing to emotional distress, anxiety, depression, loneliness, burden, and disruption to everyday life~\cite{brodaty2009family,garcia2011anxiety,manzini2020emotional,schulz2004family,harper2022,wawrziczny2017spouse,lee2022unmet}. 
Caregiving demands can also leave caregivers with limited time and capacity to attend to their own physical and mental health, relationships, and self-care~\cite{coon2009empirically,hughes2014correlates,waligora2019self,rettinger2020mixed,wawrziczny2016needing,smriti2024emotion}. 
These constraints may further limit caregivers' ability to seek formal support or maintain social engagement, particularly when daily routines become organized around the needs of the care recipient~\cite{rettinger2020mixed,wawrziczny2016needing,queluz2020understanding,smriti2024emotion}.

Importantly, caregivers' needs and their intensity are not static, but change alongside the progression of dementia~\cite{rettinger2020mixed,wawrziczny2016needing,shi2025balancing,huang2026living,kim2024opportunities}. 
As the condition progresses, caregivers may encounter new behavioral symptoms, increasing functional dependence, changing care arrangements, and greater responsibilities for decision-making~\cite{houben2024design,millenaar2018exploring,vaingankar2013perceived,shi2025balancing,wang2026taxonomy}. 
These transitions can introduce different forms of emotional and practical strain at different points in the caregiving journey~\cite{huang2026living,houben2024design,shi2025balancing,smriti2024emotion}. 
This dynamic view is consistent with the Stress Process Model, which conceptualizes caregiver wellbeing as shaped by the interaction of caregiving stressors, available resources and social support over time~\cite{pearlin1981stress}.

As \citeauthor{shi2025balancing} describe, caregivers may need to adapt to behavioral changes, navigate uncertainty around future care, or cope with changes in their relationship with the person receiving care~\cite{shi2025balancing,wawrziczny2016needing,huang2026living}. 
Prior work has therefore characterized dementia caregiving as an evolving experience in which caregivers continually adjust their roles, routines, and coping strategies in response to changing circumstances~\cite{lindgren1993caregiver,smriti2024emotion,xu2023technology,huang2026living}. 
Caregivers may postpone self-care, have difficulty accessing formal support, or rely on informal sources of assistance while managing increasingly complex responsibilities~\cite{smriti2024emotion,queluz2020understanding,waligora2019self,johnson2022s,kaliappan2025online}. 
Understanding dementia caregiving therefore requires attention not only to the care provided to the person living with dementia, but also to caregivers' changing emotional, social, and practical needs throughout the caregiving trajectory~\cite{huang2026living,shi2025balancing,steenfeldt2021becoming}.
Building on this body of work, we examine how caregiver wellbeing can be supported over time and connected, on caregivers’ terms, with the broader dementia care ecosystem.

\subsection{Digital Technologies for Caregiver Wellbeing Tracking and Reflection}

HCI and digital health research has long explored technologies that support both the practical work of caregiving and caregivers' own wellbeing~\cite{bosch2019caregiver,miller2016partners,chen2013caring,seo2019balancing,jacobs2019think,lee2023reimagining,schorch2016designing,zubatiy2021empowering,fu2025felt}. 
In dementia and older-adult care, this work spans smartphone applications~\cite{shreve2016dementia}, wearable technologies~\cite{kourtis2019digital,stavropoulos2021wearable}, voice-based systems~\cite{wong2024voice,o2020voice,pradhan2020use}, and online social platforms~\cite{johnson2022s,levonian2021patterns,pickett2024carevirtue,kaliappan2025online,saha2026ai}. 
These systems have supported care coordination, monitoring, access to information and resources, psychoeducation, social support, stress management, and other forms of self-management~\cite{cheng2024effectiveness,shin2022effects,van2017assistive,lee2022unmet,boessen2017online}. 
Systematic reviews suggest that technology-based interventions can improve outcomes such as caregiver knowledge, self-efficacy, stress, and burden, although effects vary and sustained engagement remains a challenge~\cite{cheng2024effectiveness,shin2022effects}.

Beyond functional support, research has foregrounded caregivers' emotional and psychological experiences~\cite{siddiqui2023exploring,lazar2017supporting,bhat2023we,kim2024opportunities,shi2025balancing,smriti2024emotion}. 
Prior work has examined how digital social spaces can support connection and shared coping~\cite{lazar2017supporting,johnson2022s,kaliappan2025online}, how caregivers perform emotional labor that may remain invisible within formal care~\cite{smriti2024emotion}, and how their support needs change across caregiving experiences~\cite{kim2024opportunities,shi2025balancing}. 
These concerns are particularly important because caregivers often prioritize the needs of the person receiving care over their own self-care~\cite{waligora2019self}. 
Together, this work motivates technologies that not only assist caregiving, but also create opportunities for caregivers to attend to their own wellbeing.

One approach to supporting such awareness is longitudinal tracking and reflection. 
Across HCI, self-tracking technologies have enabled people to record behaviors, experiences, and wellbeing and use those records for reflection~\cite{bhat2026my,li2010stage}. 
Repeated check-ins, self-assessments, journaling, and related forms of self-monitoring can externalize experiences that may otherwise be difficult to notice or articulate~\cite{zhang2021designing}. 
\citeauthor{epstein2015lived} conceptualize personal tracking as an ongoing process spanning collection, integration, and reflection~\cite{epstein2015lived}, while other work shows how personal data can support interpretation, self-knowledge, and recognition of relationships between experiences and wellbeing~\cite{li2010stage}. 
These capabilities may be particularly relevant for caregivers, whose attention is often directed toward monitoring another person's health rather than their own. 
Tracking centered on caregivers can instead create structured opportunities to recognize changes in their wellbeing and consider whether their support needs are being met.

More recently, conversational AI has been explored for mental health self-management and therapeutic support~\cite{sharma2024facilitating,yoo2026ai,song2025typing}, as well as to provide caregivers with accessible, on-demand wellbeing support~\cite{shi2026mapping,goel2026rubrix}.
Research on LLM-based support for dementia caregivers suggests opportunities for addressing informational and emotional needs, while also identifying limitations in contextual fit, personalization, and support for complex caregiver experiences~\cite{shi2026mapping,saha2026ai,xu2024mental}. 
Other work highlights safety risks when AI responds to caregiver distress and other high-stakes situations~\cite{goel2026rubrix}, while broader reviews document growing interest in AI-mediated support across dementia care~\cite{steijger2025use}. 
Together, this body of work motivates examining AI not as a standalone replacement for human support, but as one component of a broader caregiver wellbeing ecosystem.

\subsection{Caregiver–Clinician Communication and Collaborative Care}
Collaborative care increasingly extends beyond interactions between an individual patient and a single clinician, recognizing that care is distributed across people, settings, and forms of expertise~\cite{vizer2019s, wolff2020family,lazar2017supporting}. Contemporary collaborative care systems use patient portals, shared care plans, electronic health records, care coordination platforms, and remote monitoring technologies to support communication and information exchange across clinical and home settings~\cite{johnson2022s,levonian2021patterns,van2017assistive}. \citeauthor{chung2016boundary} highlight how these systems can help maintain continuity between clinical encounters by making information from everyday life available to care teams~\cite{chung2016boundary}. Beyond formal care systems, online communities also provide important spaces where people affected by dementia and their caregivers exchange informational and emotional support around caregiving, medical, and everyday challenges~\cite{kaliappan2025online,zhang2025mentalimager,steijger2025use}. More recently, \citeauthor{goel2026rubrix} examine AI-mediated support for dementia caregivers and highlight domain-specific risks in how AI systems respond to caregivers’ informational, emotional, and distress-related needs~\cite{goel2026rubrix}. Other work has emphasized that effective collaboration requires more than information exchange, instead depending on shared understanding and coordination among people who may hold different knowledge and responsibilities~\cite{zhu2017sharing,pollack2021different, ryu2023you,yoo2024patient,saha2026ai,steijger2025use}.

Family caregivers are important participants in these collaborative arrangements, particularly in dementia care, where they often connect everyday caregiving with formal healthcare. Caregivers observe changes in symptoms, behaviors, and functioning; coordinate appointments and services; communicate concerns to clinicians; and participate in decisions about ongoing and future care~\cite{reinhard2008supporting, given2004burden, shi2025balancing}. \citeauthor{zhai2023digital} characterize caregivers as intermediaries between patients and healthcare professionals whose experiential knowledge can provide important context beyond what is captured during clinical encounters~\cite{zhai2023digital}. Similarly, \citeauthor{gillick2013critical} shows how caregivers undertake substantial coordination work across providers and settings~\cite{gillick2013critical}. HCI research has therefore explored technologies that support caregivers in documenting observations, preparing for clinical visits, sharing information with care teams, and communicating with clinicians between visits~\cite{liang2026ll, saylor2023context,bhat2023we}. These systems position caregivers not only as recipients of clinical information, but as active collaborators in ongoing care.

However, existing collaborative care infrastructures remain largely organized around the health and care of the person receiving care, with caregivers commonly incorporated in relation to their role in supporting that care~\cite{liang2026ll,miller2016partners}. Comparatively less attention has been given to how caregivers' own wellbeing might be supported within these connected care relationships. 
This gap suggests an opportunity to move beyond individual communication tools toward a broader \textit{caregiver wellbeing ecosystem}, in which caregiver-facing technologies and clinical care can work together to support caregivers across everyday and clinical contexts. 
Such an ecosystem perspective provides a foundation for considering caregiver wellbeing as part of collaborative dementia care rather than as a separate source of support.

\section{Design Goals}
\label{sec:design-goals}

The design goals for \cmr{} were informed by prior work on caregivers' emotional labor, technology-mediated support, coordination across home and clinical settings, and changing needs throughout dementia care~\cite{smriti2024emotion,stowell2019caring,kim2024opportunities,guan2021taking,bhat2023we,shi2025balancing,shi2026mapping}. We further refined these goals through iterative discussions with our clinician coauthors.
These discussions focused on where caregiver wellbeing assessment could fit within existing neuropsychological care, what information would be useful to clinicians, and what forms of follow-up would be feasible.
The resulting goals frame \cmr{} as infrastructure for caregiver wellbeing within a broader dementia care ecosystem.
We use \emph{dementia care ecosystem} to refer to the people and services involved in dementia care, including families, clinics, and community organizations.
This framing recognizes caregivers as people with support needs of their own, rather than only as sources of information about the care recipient.

\subsection{DG1: Opening a Caregiver Wellbeing Pathway within the Dementia Care Ecosystem}
\label{sec:dg-clinical-entry}

Caregivers often accompany care recipients to neuropsychological and other dementia-related appointments.
These encounters provide an opportunity to offer caregivers support without requiring them to find and enter a separate care system.
However, caregiver participation in these visits typically centers on the care recipient's history, symptoms, and care needs.
The caregiver's own wellbeing may receive less direct attention~\cite{shi2025balancing,shi2026mapping}.

\para{DG1:} \textbf{\cmr{} should offer caregivers a brief and voluntary way to reflect on their own wellbeing during a dementia care encounter.}

To support this goal, \cmr{} invites caregivers to complete a five-to-ten-minute wellbeing assessment using established measures of caregiver wellbeing and distress~\cite{tebb2013caregiver,bangerter2019measuring}.
The wellbeing assessment is presented as separate from the care recipient's diagnostic evaluation.
Its purpose is to identify whether the caregiver may want information, follow-up, or additional support.
This point of entry is especially relevant in neuro-psychological care.
Neuro-psychological evaluations inform diagnosis and recommendations for the care recipient~\cite{weintraub2022neuropsychological}.
A caregiver wellbeing assessment can complement this process by bringing the caregiver's own needs into view.
It can also give clinicians information that may be useful when discussing the feasibility of a care plan. At the same time, incorporating validated measures of depression, anxiety, or other forms of distress into a clinical ecosystem introduces responsibilities beyond simply collecting and displaying these data. In particular, elevated scores may create an expectation that concerning results are reviewed and, when appropriate, followed by information, resources, referral, or other forms of support~\cite{brown2018caregiver}.
The caregiver wellbeing assessment is not intended to expand every clinical visit into a caregiver treatment encounter. However, this does not mean that assessment can be separated entirely from clinical responsibility once caregiver wellbeing information enters the care system. 
Instead, it creates a pathway through which caregiver needs can be acknowledged and, when appropriate, connected to follow-up support.

\subsection{DG2: Following Caregiver Wellbeing Over Time}
\label{sec:dg-longitudinal}

Caregiver wellbeing changes as symptoms, responsibilities, and family circumstances change.
A single wellbeing assessment may therefore provide only a limited view of the caregiver's experience~\cite{shi2025balancing,kim2024opportunities}.

\para{DG2:} \textbf{\cmr{} should help caregivers notice changes in their wellbeing between clinical encounters.}

The caregiver portal supports brief, repeated check-ins that allow caregivers to reflect on their wellbeing over time. 
Caregivers can review their prior responses to identify changes in their emotional state, while the timing and frequency of check-ins can be adjusted based on their preferences and recent responses. 
Drawing on prior work on timely support in digital health~\cite{nahum2016just}, \cmr{} is designed to surface changes in wellbeing without automatically inferring the mental health. 
Instead, it relies on caregiver-provided information to make patterns and shifts more visible to the caregiver. 
Beyond emotional wellbeing, check-ins can also capture the practical demands of caregiving, such as caregivers' perceived ability to manage appointments, safety concerns, or future care decisions. 
In this way, the portal recognizes caregiver wellbeing based on both emotional strain as well as ongoing responsibilities~\cite{shi2025balancing}.

\subsection{DG3: Connecting Caregivers with Timely and Personalized Support}
\label{sec:dg-personalized-support}

Caregiver needs evolve as care responsibilities, symptoms, and family circumstances change.
Support that is useful at one point may not be useful later~\cite{shi2025balancing}.
A caregiver wellbeing system should help caregivers respond when a need arises.

\para{DG3:}
\textbf{\cmr{} should help caregivers with support that reflects their current needs and circumstances.}

Based on caregivers' wellbeing assessments and check-ins, \cmr{} presents resources that are relevant to the concerns they report.
These may include self-care strategies, information about dementia care, local support groups, respite services, or opportunities to speak with a professional.
When responses indicate increasing distress, the system can surface these options without requiring the caregiver to wait until the next clinical visit.
This approach reduces the effort required to search for support across different services~\cite{guan2021taking,smriti2024emotion, kim2024opportunities}.

Recent advances in generative AI create opportunities for conversational support designed around caregivers' needs.
Accordingly, \cmr{}'s design includes a caregiver-centered chatbot that responds to concerns arising in the course of dementia caregiving.
The chatbot is aimed at providing relevant information, supporting structured reflection, and suggesting coping strategies based on the concerns caregivers describe~\cite{shi2026mapping}.
It can also direct caregivers toward professional or community resources when additional support may be appropriate.
The chatbot complements the other features of \cmr{} by offering another way to access support between clinical encounters.
It is not intended to provide clinical care, make diagnoses, or replace support from a qualified professional.

\subsection{DG4: Supporting Clinician Awareness and Caregiver Follow-Up}
\label{sec:dg-clinician-followup}

Clinicians may have limited information about caregiver wellbeing between clinical encounters.
Caregivers also frequently coordinate information and care across home and clinical settings~\cite{bhat2023we}.
Without a structured way to communicate caregiver needs, signs of distress may remain outside the clinical conversation.

\para{DG4:}
\textbf{\cmr{} should help clinicians understand caregiver-reported needs and determine when follow-up may be appropriate.}

The clinician-facing dashboard presents caregiver wellbeing assessment results, changes in wellbeing, and requests for support.
It also provides information about caregivers' engagement with recommended resources.
This view allows clinicians to identify patterns without reviewing every caregiver interaction.

Clinicians can use this information to tailor care discussions, recommend services, and follow up with caregivers who may need additional support.
The information may also help clinicians consider whether a care plan places demands on the caregiver that are difficult to sustain.
Bringing caregiver wellbeing into the clinical conversation can therefore support both the caregiver and the care recipient.

The dashboard is intended to inform clinical judgment rather than automate it.
It does not diagnose caregivers or determine what action a clinician must take.
Clinicians decide how to interpret the information and whether follow-up is needed.
Caregiver control over information sharing is also central to this goal.
Caregivers may want clinicians to understand their wellbeing without sharing every response or conversation.
The evaluation therefore examines what information caregivers are comfortable sharing and how they expect clinicians to respond.

\section{Designing \cmr{}} \label{sec:system}

We designed a caregiver--clinician ecosystem that bridges caregivers' everyday wellbeing with clinical attention and care. The \cmr{} ecosystem consists of two interconnected interfaces: a caregiver-facing portal that supports longitudinal wellbeing tracking and reflection, and a clinician-facing portal that enables clinicians to understand caregivers' wellbeing over time. Together, these components explore how caregiver wellbeing information can move between a caregiver's lived experience and clinical contexts while preserving appropriate boundaries regarding privacy and data access. In this section, we describe the design process and clinical context of \cmr{}, both its caregiver- and clinician-facing components, and the technical architecture that connects the two sides of the ecosystem. 

\subsection{Design Process and Clinical Context}
The design of \cmr{} was influenced by prior research examining the wellbeing needs of caregivers of people with dementia~\cite{brodaty2009family, steenfeldt2021becoming, shi2025balancing}. This work highlighted that caregivers often experience emotional and practical challenges, yet support for their own wellbeing remains largely absent from clinical care centered on the person with dementia~\cite{manzini2020emotional, seidel2019burden, yuan2025supporting, shi2025balancing}. Motivated by this gap, we designed \cmr{} to explore how caregivers' everyday wellbeing experiences could become more visible and actionable for both caregivers and clinicians.

\begin{figure*}[t]
    \centering
    \includegraphics[width=0.9\textwidth]{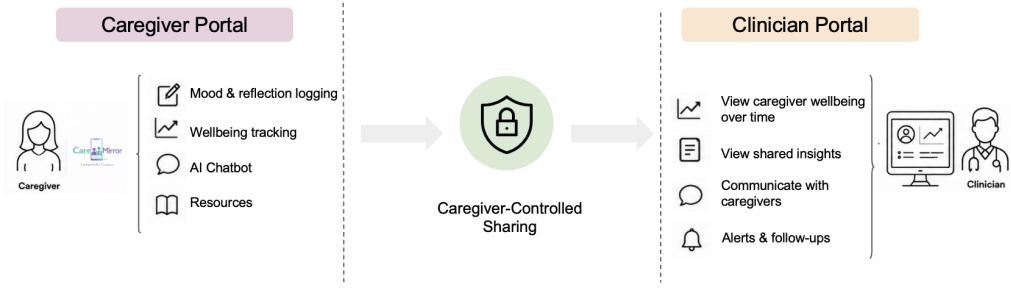}
     \caption{Overview of the \cmr{} caregiver-clinician ecosystem. \cmr{} supports caregivers in tracking and reflecting on their wellbeing while connecting them with clinical care through caregiver-controlled sharing. Caregivers determine what information is shared, while clinicians can review wellbeing information, communicated with caregivers, and provide follow-up support. }
     \Description{Overview of the \cmr{} caregiver--clinician ecosystem. The caregiver-facing side supports wellbeing tracking, reflection, and personalized support, while the clinician-facing side supports review of caregiver wellbeing information and follow-up. The two sides are connected through caregiver-controlled sharing, in which caregivers determine what wellbeing data and insights are shared with clinicians.}
    \label{fig:ecosystem}
\end{figure*}

\autoref{fig:ecosystem} illustrates this \cmr{} ecosystem and how caregivers' everyday wellbeing can connect with clinical care. On the caregiver side, \cmr{} provides mechanisms for caregivers to regularly document their wellbeing, reflect on changes over time, and chat with an AI chatbot and resources. On the clinician side, a complementary portal provides a longitudinal view of caregiver wellbeing information that has been made available for clinical review. The two interfaces are connected through caregiver-controlled sharing, allowing wellbeing information to move from caregivers' everyday experiences towards clinical awareness and follow-up while preserving caregivers' control over what information enters the clinical context.

\subsection{Caregiver-Facing Portal}

The caregiver-facing portal serves as the primary interface through which caregivers interact with \cmr{}. 
\cmr{} was originally developed in Figma and iteratively refined over several weeks an author in collaboration with the broader research team. Clinician coauthors were also closely involved in the design process, particularly in shaping how caregiver wellbeing information could connect with the clinician-facing portal.

\begin{figure*}[t]
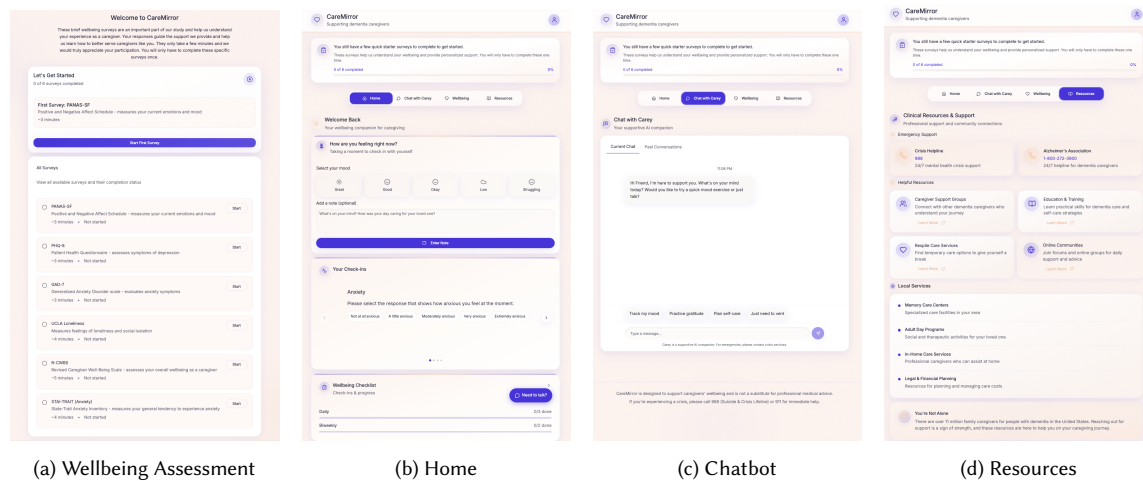

    \centering

    \begin{subfigure}[t]{0.235\textwidth}
        \centering
        \includegraphics[width=\linewidth]{_caregiver_survey.pdf}
        \caption{Wellbeing Assessment}
        \label{fig:caregiver-survey}
    \end{subfigure}
    \hfill
    \begin{subfigure}[t]{0.235\textwidth}
        \centering
        \includegraphics[width=\linewidth]{_caregiver_home.pdf}
        \caption{Home}
        \label{fig:caregiver-home}
    \end{subfigure}
    \hfill
    \begin{subfigure}[t]{0.235\textwidth}
        \centering
        \includegraphics[width=\linewidth]{_caregiver_carey.pdf}
        \caption{Chatbot}
        \label{fig:caregiver-chat}
    \end{subfigure}
    \hfill
    \begin{subfigure}[t]{0.235\textwidth}
        \centering
        \includegraphics[width=\linewidth]{_caregiver_resource.pdf}
        \caption{Resources}
        \label{fig:caregiver-resources}
    \end{subfigure}

    \caption{\cmr{} caregiver-facing interface: 
    (a) The wellbeing assessment supports baseline and ongoing assessment of caregiver wellbeing.
    (b) The home page supports mood check-ins and longitudinal wellbeing tracking.
    (c) The chatbot provides a space for caregiver reflection and support.
    (d) The resources page provides access to clinical, community, and local support options.}
    \Description{Four screenshots of the \cmr{} caregiver-facing interface. Panel (a) shows wellbeing assessments for reporting aspects of caregiver mental and emotional wellbeing. Panel (b) shows the home page with a mood check-in, wellbeing information, and daily and biweekly assessments. Panel (c) shows the AI chatbot for caregiver reflection, questions, and support. Panel (d) shows clinical, community, educational, respite, emergency, and local caregiver support resources.}
    \label{fig:caregiver-portal}
\end{figure*}

When caregivers first use \cmr{}, they are invited to complete baseline wellbeing assessment: (PANAS-SF~\cite{thompson2007development}, PHQ-8~\cite{kroenke2009phq}, GAD-7~\cite{spitzer2006brief}, UCLA Loneliness~\cite{russell1996ucla}, R-CWBS~\cite{tebb2013caregiver}, and STAI-TRAIT (Anxiety)~\cite{renzi1985state}) (Figure~\ref{fig:caregiver-survey}). Caregivers can start, save, skip, or return to these wellbeing assessments later. To support longitudinal tracking while reducing burden, \cmr{} subsequently uses shorter biweekly measures (GAD-2~\cite{kroenke2007anxiety}, PHQ-2~\cite{kroenke2003patient}) and daily measures (PANAS-Short~\cite{thompson2007development}). Wellbeing assessment results are presented longitudinally, with indicators of potential concern made available for clinician review rather than treated as diagnoses or automatically triggering intervention.


The home page provides a lightweight mood check-in that asks caregivers, ``How are you feeling right now?'' Caregivers can select from five moods, optionally add a note, and complete their daily or biweekly check-ins (\autoref{fig:caregiver-home}). These responses contribute to longitudinal wellbeing trends and, with caregiver permission, can inform the clinician-facing portal (\autoref{fig:clinician-detail}).

The caregiver-centered AI chatbot provides a space for caregivers to reflect on their experiences and seek support (\autoref{fig:caregiver-chat}). Conversation starters facilitate interaction, 
while past conversations allow caregivers to revisit previous chat history. The interface includes a disclaimer distinguishing \cmr{} from clinical support and directing caregivers to crisis services in emergencies. 

Finally, the resources page brings together clinical, community, and local support options (\autoref{fig:caregiver-resources}), including emergency support, caregiver support groups, education and training, respite care, online communities, and local care services. Together, these components connect ongoing wellbeing tracking with opportunities for reflection and support within a single caregiver-facing interface.

\subsection{Clinician-Facing Portal}
To complement the \cmr{}, we designed a clinician-facing dashboard that centralizes caregiver wellbeing information and supports clinicians in identifying caregivers who may need additional attention. The dashboard provides population-level and individual-level views, allowing clinicians to move from an overview of their caregiver caseload to more detailed information about a specific caregiver.

\begin{figure*}[t]
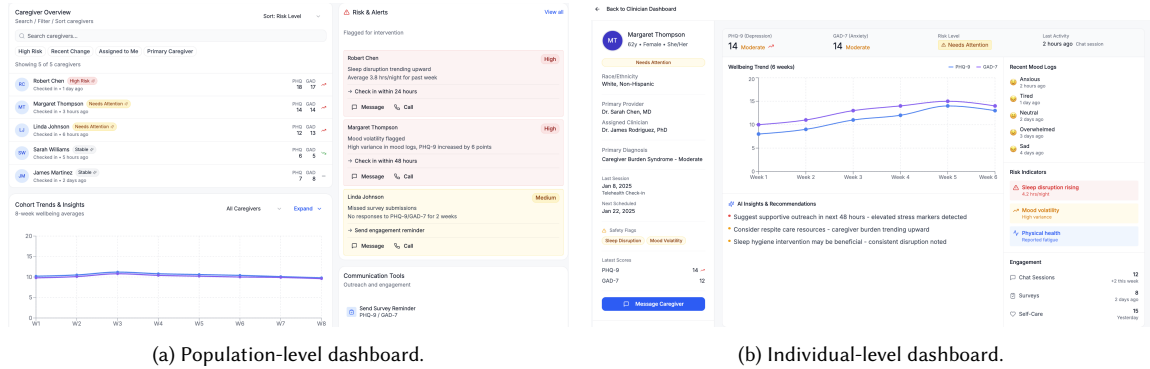

    \centering

    \begin{subfigure}[t]{0.49\textwidth}
        \centering
        \includegraphics[width=\columnwidth]{_clinician_overview.pdf}
        \caption{Population-level dashboard.}
        \label{fig:clinician-dashboard}
    \end{subfigure}
    \hfill
    \begin{subfigure}[t]{0.49\textwidth}
        \centering
        \includegraphics[width=\columnwidth]{_clinician_individual.pdf}
        \caption{Individual-level dashboard.}
        \label{fig:clinician-detail}
    \end{subfigure}

    \caption{\cmr{} clinician-facing portal. 
    (a) The population-level dashboard provides an overview of caregiver wellbeing across the clinician's caseload. (b) The individual-level dashboard presents a caregiver's longitudinal wellbeing information, AI-generated insights, and options for clinician follow-up.}
    \Description{Two screenshots of the \cmr{} clinician-facing portal. Panel (a) shows a population-level dashboard with caregivers, indicators of potential concern, recent check-in activity, wellbeing measures, and aggregate trends. Panel (b) shows an individual caregiver dashboard with longitudinal wellbeing trends, assessment information, AI-generated insights and recommendations, indicators of potential concern, and options 

for clinician follow-up.}

    \label{fig:clinician-portal}
  
\end{figure*}

The population-level dashboard provides an overview of caregiver wellbeing across a clinician's caseload (~\autoref{fig:clinician-dashboard}). Clinicians can review recent check-in activity, wellbeing measures, and indicators of potential concern, as well as broader wellbeing and assessment completion trends. These indicators are intended to direct clinician attention to patterns that may warrant further review rather independently characterize a caregiver's clinical or safety risk.

Clinicians can select a caregiver to access an individual-level dashboard that brings together longitudinal wellbeing trends, indicators of potential concerns, AI-generated insights, and \cmr{} activity (Figure~\ref{fig:clinician-detail}). AI-generated insights help clinicians identify potentially relevant patterns while supporting, rather than replacing, clinician review of caregiver-reported information. The interface also provides pathways for follow-up, including caregiver outreach, assessment reminders, and session scheduling.

\begin{figure*}[t]
    \centering
    \includegraphics[width=0.95\textwidth]{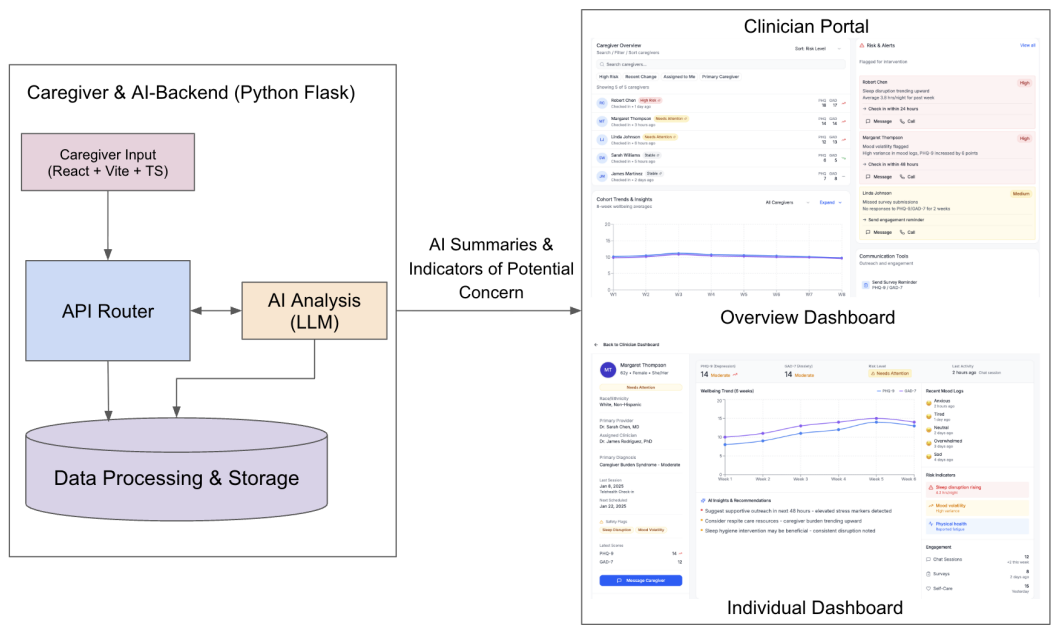}
    \caption{Technical architecture and data flow of \cmr{}. Caregiver-generated wellbeing data are routed through the Flask backend for processing and storage and analyzed by the LLM-based component to generate wellbeing summaries and indicators of potential concern. These outputs are surfaced through the clinician portal at both overview and individual-caregiver levels to support asynchronous clinical awareness and review.
    }
    \Description{Technical architecture and data flow of \cmr{}. Caregiver-generated wellbeing information enters through a React, Vite, and TypeScript caregiver interface and is routed through a Python Flask API. Data are processed and stored longitudinally and can be analyzed by an LLM-based component. The analysis generates wellbeing summaries and indicators of potential concern, which are presented through population-level and individual-level dashboards in the clinician portal.}
    \label{fig:technical_architecture}
    
\end{figure*}
\subsection{System Architecture and Data Flow}

\autoref{fig:technical_architecture} shows the technical architecture and data flow of \cmr{}. The caregiver-facing interface was implemented using React, Vite, and TypeScript, with a Python Flask backend connecting the interface, data storage, and AI components.
Caregiver-generated data, including mood check-ins, wellbeing assessments, and AI chatbot interactions, are processed and stored to maintain a longitudinal record of caregiver wellbeing. 
The backend also connects these longitudinal data with an LLM-based analysis component to generated wellbeing insights and indicators of potential concern. These outputs are surfaced through the clinician portal at both population- and individual- level to support clinical awareness and review. Indicators of potential concern are intended to direct clinician attention raster than function as real-time alerts or replace clinician review of the underlying caregiver-reported information.

The envisioned use of \cmr{} within a healthcare ecosystem introduces additional clinical safety considerations, around crisis disclosures, wellbeing assessments, and indicators of potential concerns. \cmr{} is designed to support asynchronous clinical awareness rather than real-time monitoring and should not imply immediate clinician response. 
Future deployment would require protocols for reviewing concerning information, follow-up and escalation, and communicating these expectations to caregivers. These workflows were not implemented or evaluated in the present study.

\section{User Study}
We conducted a user study with caregivers to understand how they envision proactive technologies supporting their own wellbeing and how this support might connect with the broader dementia care ecosystem. Using \cmr{} as a design probe~\cite{hutchinson2003technology}, we explored caregivers’ perceptions of the proposed ecosystem and surfaced their expectations, concerns, and boundaries around its use and integration with clinical care. This section describes our recruitment, study procedure, and data analysis.
\subsection{Participant Recruitment}
We recruited participants primarily through social media and online communities related to dementia and caregiving. We posted recruitment information on Reddit communities (e.g., r/dementia and r/dementiaresearch). Recruitment posts directed interested individuals to an interest form, which served as an initial screening tool to assess eligibility. Participants were required to (1) be 18 years of age or older, (2) be a current or former caregiver for a person living with dementia, and (3) currently reside in the U.S. We received 169 responses to our interest form over approximately five months from March 2026 to August 2026. Eligible respondents were contacted for a brief phone screening to confirm eligibility and study fit before being invited to participate. Ultimately, we interviewed 14 caregivers, including 11 family caregivers and 3 professional caregivers. Each participant received a \$25 Amazon gift card as compensation. \autoref{tab:participants} summarizes the demographic characteristics of our participant pool.

\begin{table}[t]
\centering
\sffamily
\footnotesize
   \caption{Summary of participants, including type (current/former caregiver), years of caregiving (ys.), care-recipient, demographics, technology used for mental wellbeing (if any), and employment status. \textit{Professional Caregivers} are marked with an `*' next to their ID.}
   \label{tab:participants}
\setlength{\tabcolsep}{3pt}
\resizebox{\columnwidth}{!}{
\begin{tabular}{llrp{0.1\columnwidth}rlp{0.08\columnwidth}lp{0.1\columnwidth}l}
\textbf{ID} & \textbf{Type} & \textbf{Ys.} & \textbf{Care Recipient} & \textbf{Age} & \textbf{Sex} & \textbf{Race} & \textbf{Education} & \textbf{Tech. Used} & \textbf{Employment} \\
\toprule
P1  & Former  & $<$1 & Mother     & 50-65  & Female      & White                      & Advanced degree                    & \edit{Social media, smartphone}                                           & Retired \\
\rowcollight P2  & Former     & 2  & Father     & 50-65  & Female      & White                      & Advanced degree                    & \edit{Smartphone, social media, computer, ChatGPT}                        & Self-employed \\
P3  & Current & 7  & Stepfather & 25-35  & Female      & Black or African American  & Some college, no degree            & \edit{Smartphone and social media}                                        & Self-employed \\
\rowcollight P4  & Former  & 2  & Parents    & 36-50  & Female      & White                      & Advanced degree                    & \edit{Zoom (therapy), Reddit, home automation tech}                       & Employed for wages \\
P5  & Current & 8  & Mother     & 50-65  & Female      & White                      & Associate degree                   & \edit{---}                                                                & Out of work \\
\rowcollight P6  & Current & 3  & Father     & 25-35  & Male        & Black or African American  & Bachelor's degree                  & \edit{Meditation/journaling apps, Reddit, Facebook groups, online therapy} & Employed for wages \\
P7  & Current & 5  & Care recipient (client)       & 25-35  & Male        & White                      & Bachelor's degree                  & \edit{Laptop}                                                             & Self-employed \\
\rowcollight P8  & Former & 18 & Mother-in-law, grandmother, uncle & 36-50 & Female & Asian & Advanced degree           & \edit{Smartphone}                                                         & Employed for wages \\
P9  & Former & 10+ & Husband    & 50-65  & Female      & White                      & Advanced degree                    & \edit{Smartphone}                                                         & Retired \\
\rowcollight P10 & Former & 5  & Father-in-law & 50-65 & Female     & Asian                      & Advanced degree                    & \edit{---}                                                                & Self-employed \\
P11* & Former & 5+ & Care recipients (clients)     & 25-35  & Female      & Asian                      & Advanced degree                    & \edit{Remote counseling sessions}                                         & A student \\
\rowcollight P12* & Current & 2  & Resident at memory care center & 19-24 & Female & White                & Some college, no degree            & \edit{---}                                                                & Employed for wages \\
P13* & Former & 4  & Residents (former employer, assisted living facility) & 25-35 & Male & White       & Bachelor's degree                  & \edit{Online therapy, meditation apps}                                    & Employed for wages \\
\rowcollight P14 & Current & 18 & Wife       & 66+    & Male        & White                      & Advanced degree                    & \edit{---}                                                                & Retired \\
\bottomrule
\end{tabular}}
\Description[table]{Characteristics of the 14 study participants. Columns report participant ID, current or former caregiver status, years of caregiving, relationship to the care recipient, age range, sex, race, education, technologies used for mental wellbeing, and employment status. Three participants are marked as having professional caregiving experience.}
\end{table}

\subsection{Interview Procedure}
We conducted semi-structured interviews with caregivers to explore their caregiving experiences, wellbeing needs, and perspectives on using AI-assisted system for caregiver support. Interviews were conducted remotely via Zoom, lasted approximately 60 minutes and were recorded with participants' consent. The interview protocol consisted of four stages: (1) understanding participants' caregiving experiences and wellbeing needs, (2) guided interaction with \cmr{}, (3) post-interaction reflection on \cmr{} and future caregiving wellbeing support, and (4) discussion of a caregiver--clinician wellbeing ecosystem, including clinical connection, information sharing, privacy and control, and the role of AI in caregiving. 

We began by asking participants about their caregiving responsibilities, how their wellbeing had changed throughout their caregiving journey, and their experiences with existing sources of support. Participants were then introduced to \cmr{} and guided through its key components, including mood check-ins, wellbeing assessments, the AI chatbot, and caregiver resources. We encouraged participants to think aloud while interacting with the system and to describe what they found useful, what they would change, and how these features might fit into their everyday caregiving routines. Rather than evaluating task completion, we used \cmr{} as a design probe~\cite{hutchinson2003technology} to elicit participants' needs, expectations, and preferences.

After interacting with \cmr{}, participants reflected on their overall experience and how they envisioned such support evolving over time. We then introduced the broader caregiver--clinician ecosystem and asked participants to consider how their wellbeing information might connect with clinical care. We discussed what information they would or would not want to share, who should have access, how sharing should be controlled, and their expectations and concerns regarding AI-generated interpretations of their wellbeing.

\subsection{Data Analysis}
\subsubsection{Descriptive Analysis}
Our descriptive analysis drew on the entry and exit surveys completed by all $N=14$ participants. At entry, participants completed the Rapid-Caregivers' Wellbeing Scale (R-CWBS; \cite{tebb2013caregiver}) and rated a set of prompts on caregiver mental wellbeing concerns drawn from the literature, both on 5-point scales. 
\autoref{table:cwbs} and \autoref{table:mentalwellbeing} summarize participants' entry survey responses to the R-CWBS and measures of caregiver mental wellbeing concerns, respectively---showing a range of caregiving-related wellbeing challenges and responsibilities represented in our participant pool.
At exit, participants evaluated \cmr{} using the System Usability Scale (SUS) \cite{brooke1996sus} and the Intervention Appropriateness Measure (IAM) \cite{weiner2017psychometric}. 
For all entry- and exit-survey items, we computed descriptive statistics (mean, median, standard deviation), following standard SUS and IAM scoring conventions.

\begin{table}[t!]
\centering
\sffamily
\footnotesize
\caption{Summary of participants' responses to Rapid-Caregivers' Wellbeing Scale (R-CWBS)~\cite{tebb2013caregiver}. Each question was rated on: 1 (Rarely), 2 (Occassionally), 3 (Sometimes), 4 (Usually), and 5 (Frequently).}
\label{table:cwbs}

\begin{tabular}{lrrrc}
\textbf{Question} & \textbf{Mean} & \textbf{\edit{Median}} & \textbf{Std. Dev.} & \textbf{Distribution}\\
\toprule
\rowcollight \multicolumn{5}{c}{\textbf{Activities}}\\
Taking care of personal daily activities (meals, hygiene, laundry) & 4.00 & \edit{4} & 0.78 & \includegraphics[height=6pt]{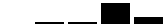}\\
Taking time to have fun with friends and/or family & 2.86 & \edit{3} & 1.29 & \includegraphics[height=6pt]{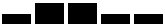}\\
Treating or rewarding yourself & 2.86 & \edit{3} & 1.35 & \includegraphics[height=6pt]{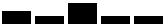}\\
\rowcollight \multicolumn{5}{c}{\textbf{Needs}}\\
Receiving appropriate health care & 3.71 & \edit{4} & 1.33 & \includegraphics[height=6pt]{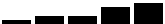}\\
Feeling good about yourself & 3.29 & \edit{3} & 1.33 & \includegraphics[height=6pt]{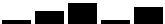}\\
Feeling secure about your financial future & 3.00 & \edit{3} & 1.47 & \includegraphics[height=6pt]{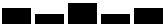}\\
\bottomrule
\Description[]{Summary of participants' responses to six items from the Rapid-Caregivers' Wellbeing Scale, rated from 1, rarely, to 5, frequently. The table reports the mean, median, standard deviation, and response distribution for three activity items and three needs items. Mean scores range from 2.86 for having fun with friends or family and treating or rewarding oneself to 4.00 for taking care of personal daily activities.}
\end{tabular}

\end{table}

\begin{table}[t!]
\centering
\sffamily
\footnotesize
\caption{Summary of participants' responses to prompts on mental wellbeing concerns drawn on the literature~\cite{shi2025balancing}. Participants responded to these prompts based on how much they associated with these concerns, on a scale of 1 (not at all concerning) to 5 (very concerning).}
\label{table:mentalwellbeing}
\begin{tabular}{lrrrc}
\textbf{Question} & \textbf{Mean} & \textbf{\edit{Median}} & \textbf{Std. Dev.} & \textbf{Distribution}\\
\toprule
Disruptive Behaviors by Care Recipients & 4.29 & \edit{5} & 1.27 & \includegraphics[height=8pt]{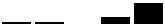}\\
\rowcollight Lack of Support & 3.86 & \edit{4} & 0.95 & \includegraphics[height=6pt]{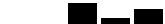}\\
Low Self-Efficacy & 3.29 & \edit{3} & 1.07 & \includegraphics[height=6pt]{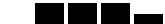}\\
\rowcollight Emotional Distress & 3.86 &\edit{4} & 1.10 & \includegraphics[height=6pt]{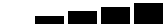}\\
Relationship Tensions & 3.93 & \edit{4} & 0.92 & \includegraphics[height=6pt]{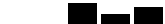}\\
\rowcollight Compassion Fatigue & 4.00 & \edit{4} & 0.96 & \includegraphics[height=6pt]{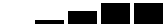}\\
Lack of Self-Care & 3.71 & \edit{4} &  1.20 & \includegraphics[height=6pt]{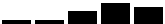}\\
\rowcollight Burnout & 4.07 & \edit{4} & 1.00 & \includegraphics[height=6pt]{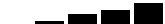}\\
\bottomrule
\Description[table]{Summary of participants' ratings of eight caregiver mental wellbeing concerns on a scale from 1, not at all concerning, to 5, very concerning. The table reports the mean, median, standard deviation, and response distribution for disruptive behaviors, lack of support, low self-efficacy, emotional distress, relationship tensions, compassion fatigue, lack of self-care, and burnout. Disruptive behaviors have the highest mean rating at 4.29, while low self-efficacy has the lowest at 3.29.}

\end{tabular}
\end{table}

\subsubsection{Qualitative Analysis}

Following the interviews, all recordings were transcribed using Zoom transcription feature. We anonymized the transcripts by removing identifiable information and analyzed the transcripts alongside interview notes. We conducted reflexive thematic analysis~\cite{braun2019reflecting} through an iterative process of open coding. Two authors led the initial coding of the interview transcripts, while the broader research team regularly reviewed and discussed emerging codes and interpretations during collaborative co-working sessions. The two authors subsequently synthesized the initial codes into higher-level themes, with the broader team providing feedback throughout the process.
Theme development was iterative and guided by our research questions. We repeatedly revisited the transcripts, codes, and emerging themes, grouping conceptually related codes, distinguishing themes that captured different aspects of caregivers' experiences, and setting aside codes that were less relevant to our research questions. Through this process, we refined an initial set of 467 open codes into 4 high-level themes capturing caregivers' expectations, concerns, and boundaries around proactive wellbeing support and its connection to clinical care.

\subsection{Privacy, Ethics, and Reflexivity}
Our study was reviewed and approved by the Institutional Review Boards (IRBs) at our institution. Given the sensitive nature of discussing caregiving and personal wellbeing, we adopted several measures to protect participants' privacy and emotional wellbeing. Participants were assigned unique IDs, identifying information was removed from transcripts prior to analysis, and study data were stored in secure, access-controlled environments. During interviews, participants could decline to answer questions, pause, or discontinue the session at any time, and interviewers remained attentive to signs of discomfort when discussing emotionally challenging caregiving experiences. Because \cmr{} explores sharing caregiver wellbeing information with clinical care teams, we explicitly invited participants to discuss what information they would or would not want to share, with whom, and under what circumstances, treating these boundaries as central to our inquiry. Our interdisciplinary research team brings expertise in HCI, human-centered AI, digital mental health, and clinicians. 
Among the clinician coauthors, one specializes in clinical psychology with over 16 years of experience in adult and adolescent inpatient care and crisis suicide helpline services, while another specializes in neuropsychology and is an active clinical practitioner working with individuals living with dementia and their caregivers.
We acknowledge that these perspectives may shape our interpretations and therefore engaged in collaborative and iterative discussions throughout the analysis to remain attentive to participants' perspectives, including those that challenged assumptions embedded in the design of \cmr{}.

\section{Findings}
Overall, participants welcomed the \cmr{}'s potential to extend support beyond one-time wellbeing assessments through longitudinal tracking, AI-supported reflection, and connection with clinical care. Participants were generally receptive to these forms of support, particularly when \cmr{} could help them recognize changes in their wellbeing over time and reduce the burden of communicating these changes to clinicians. 
At the same time, participants raised important concerns about privacy, data sharing, and the extent to which AI and clinicians should have access to their personal wellbeing information. 
These considerations highlight that connecting caregiver wellbeing with clinical care requires not only technical integration, but also mechanisms that preserve caregivers' agency and control.

\subsection{RQ1: Caregivers' Perceptions about \cmr{}}\label{sec:caregiver-response}

Participants' interactions with \cmr{} provided insight into how they understood and experienced the system.
First, we report results from the post-interaction surveys assessing perceived usability and appropriateness. 
Then, we organize participants' qualitative insights around the four design goals (DGs), highlighting both the value participants identified in the current design and the tensions that emerged through their responses.

\subsubsection{Perceived Usability and Appropriateness}
\label{sec:6.1.1}
\begin{figure}[t]
\centering
\includegraphics[width=0.8\linewidth]{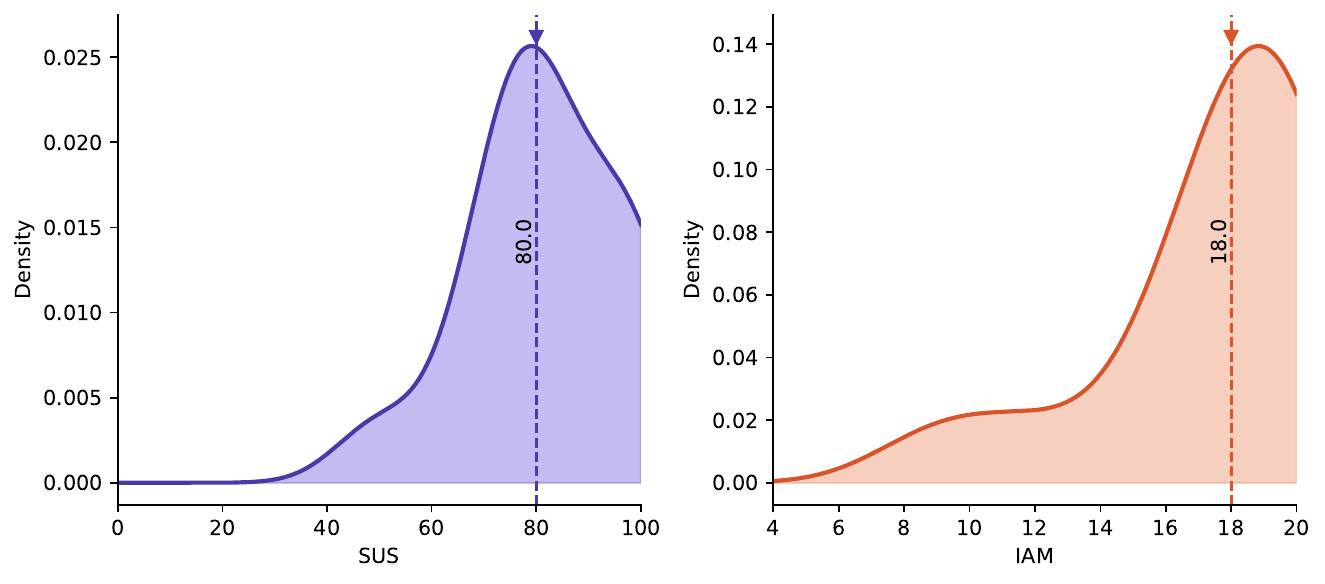}
\caption{Distribution plots of SUS and IAM scores based on the exit surveys (the dotted lines represent the median values of the corresponding color distribution).}
\Description{
Distribution plots of System Usability Scale (SUS) and Intervention Appropriateness Measure (IAM) scores from the exit survey. The SUS scores have a mean of 81.61 and median of 80 on a 0--100 scale. The IAM scores have a mean of 17.43 and median of 18 on a 4--20 scale. Dotted lines mark the median values of the corresponding distributions.}
\label{fig:sus_iam}
\end{figure}

During the exit survey, participants completed the System Usability Scale (SUS; 0--100)~\cite{brooke1996sus} and the Intervention Appropriateness Measure (IAM; 4--20)~\cite{weiner2017psychometric} to evaluate their experience in terms of usability and appropriateness with \cmr{}. 
The distribution plots are shown in~\autoref{fig:sus_iam}. 
The mean SUS score was 81.61 (median=80.0), which, consistent with prior work, is considered above average and acceptable usability~\cite{brooke1996sus}. 
The mean IAM score was 17.43 (median=18.0), indicating that participants perceived the intervention as highly appropriate~\cite{weiner2017psychometric}. 
This suggests that within our limited participant pool, \cmr{} was viewed as a strong fit with clear value for supporting caregivers' digital wellbeing.

Participants identified value in \cmr{} for supporting ongoing self-reflection, particularly through the ability to revisit previous experiences and reflect on changes in their wellbeing over time. As P6 described: 
\begin{quote}
\small
``I think it's helpful to see my past conversations. If I had vented about something and it helped, I would love to go back and see what I was upset about and what helped me at the time.'' (P6)
\end{quote}

That said, we note that \cmr{} was only explored as a prototype rather than a fully functional system, participants also commented on the limited fidelity of some interface components, which may have influenced their usability ratings. 
For example, one participant described the mood check-in interface as visually basic:

\begin{quote}
\small
``Doesn't really give me an emotional response one way or another. I know that it's automated, and that it's part of the HTML code, or whatever the code is behind the page.'' (P1)
\end{quote}


\subsubsection{Clinical Encounters Could Provide an Entry Point for Caregiver Support}
\label{sec:6.1.2}
Participants generally responded positively \cmr{}'s explicit attention to caregiver wellbeing. 
Rather than focusing only on the care recipient, the wellbeing assessment created an opportunity for caregivers to reflect on emotional and mental health concerns that might otherwise receive limited attention within dementia care. 
P8, for example, responded positively to the range of wellbeing domains included in the assessment:

\begin{quote}
\small
``The components you have listed---emotions, mood, loneliness, depression---are all helpful for caregivers, because they focus on how the caregiver is doing.'' (P8)
\end{quote}

Participants also described a broader lack of support for caregivers themselves, suggesting why such an entry point could be valuable. P10 described caregivers as needing support not only how to provide dementia care, but also in managing their own emotional experience:
\begin{quote}
\small
``Caregivers need to be literally trained on this, emotionally, how to handle themselves, keep their lives intact.'' (P10)
\end{quote}

That said, participants also cautioned that introducing caregiver wellbeing assessment into an already demanding caregiving routine could create additional burden. 
While P8 appreciated the attention to caregiver wellbeing, they also described repeated wellbeing
activities as another task to complete, noting, \textit{``I feel like [it's] a to-do task for me.''} 
This tension suggests that creating an entry point for caregiver support should not introduce another obligation into caregivers' routines, and reinforces DG's emphasis on keeping wellbeing reflection brief and voluntary. 
Beyond the effort required to complete repeated assessments, longitudinal symptom monitoring may itself carry emotional consequences. While regularly reflecting on wellbeing may help some caregivers recognize changes over time, repeatedly directing attention toward distress may feel burdensome or heighten symptom-focused attention for others. Although our study did not evaluate these effects through longitudinal use of \cmr{}, participants’ concerns about repeated assessment suggest that the frequency and timing of wellbeing check-ins should be configurable rather than assumed to be uniformly beneficial.

Together, participants saw value in creating an explicit space for caregiver wellbeing within the broader context of dementia care, particularly when it brought emotional needs into view that might otherwise receive limited attention. At the same time, this pathway should minimize additional demands on caregivers, allowing wellbeing assessment to function as an accessible opportunity for support rather than another caregiving responsibility.



\subsubsection{Longitudinal Awareness Was Valuable but Could Also Reinforce Distress}
\label{sec:6.1.3}
Participants valued \cmr{} not simply as a tool for recording how they felt at a particular moment, but as a way to develop a longitudinal understanding of their wellbeing. 
Participants described caregiver wellbeing as highly dynamic and closely intertwined with changing caregiving demands, making isolated check-ins difficult to interpret without temporal context. 
For instance, P1 described caregiver moods as fluctuating substantially over time.

\begin{quote}
\small
``Caregiver moods are really kind of yo-yos, so having a time frame for the questions---whether it's the last hour, day, or week---would have helped me answer them.'' (P1)
\end{quote}

Beyond individual check-ins, participants also saw value in \cmr{}'s ability to preserve previous mood entries and associated reflections. 
P1 described how returning to notes from a better day could provide reassurance when reflecting on their wellbeing: \textit{``I'd be able to go back to a day when I felt great or good, and use that to reassure myself. So, I like having the notes.'' (P1)} 
These responses highlight how \cmr{}'s longitudinal features could extend the value of individual check-ins by allowing caregivers to revisit previous entries and place their current wellbeing in relation to earlier experiences. 
At the same time, participants noted that preserving and resurfacing this history was not always beneficial. 
P1 explained that the emotional effect of revisiting previous records could depend on both the content of the record and their current emotional state. 
In particular, encountering a record of a difficult caregiving experience while currently feeling well could itself become distressing:

\begin{quote}
\small
``I might turn to \cmr{} only when I'm feeling overwhelmed and ignore it when I'm actually feeling pretty good. So, if I had three crises over the last three days, having it tell me that it thinks I'm in crisis could trigger that kind of spiraling.'' (P1)
\end{quote}

These responses both supported and complicated DG2. 
While \cmr{}'s longitudinal features could help caregivers revisit previous experiences and contextualize changes in their wellbeing, the same features could become emotionally burdensome when they resurfaced difficult moments. 
Participants' feedback therefore highlights an important consideration for refining \cmr{}'s longitudinal views: preserving the reflective value of wellbeing history while remaining sensitive to when and how distressing past experiences are brought back into view.

\subsubsection{Timely Support Needed to Reflect Each Caregiver's Context}
\label{sec:6.1.4}

Participants generally saw value in \cmr{}'s use of AI to provide support based on their current wellbeing and caregiving experiences~\cite{shi2026mapping}. 
Participants responded positively to the AI chatbot as an immediately available source of guidance and to \cmr{}'s resources as a way of connecting their reported wellbeing with relevant support. 
Across their responses, the usefulness of these features was closely tied to how well the support reflected the caregiver's particular circumstances.

Participants' interactions with the chatbot highlighted the importance of this contextual fit. 
P7 described the chatbot's responses as helpful in providing general guidance, while also noting that such guidance did not always capture the particular person or caregiving situation involved:

\begin{quote}
\small
``The responses are good, they're guiding, [but] they give you general [advice] ... it cannot fit the personality.'' (P7)
\end{quote}
The degree of personalization also shaped how participants experienced their interactions with the chatbot. P10 described the conversational nature of the chatbot as contributing to a more individualized experience: \textit{``That can give you a much more personal experience.''} 
Participants also questioned whether AI-generated support could provide the same quality of response as human support. P2 evaluated the chatbot's response as adequate while still identifying a gap between the guidance provided by AI and what they might expected from a person:
\begin{quote}
\small
``The response wasn't quite as good as what I might get from a human. I thought it was \dots fair.'' (P2)
\end{quote}
These concerns suggest that immediate availability alone did not make AI support personally meaningful. Participants evaluated the chatbot not only by whether it could provide guidance, but also by whether that guidance reflected their individual circumstances and approached the quality of support they associated with human interaction. 

A similar emphasis on contextual relevance emerged in participants' responses to \cmr{}'s resource recommendations. 
P1 reflected on how resources associated with different emotional experiences could become useful when they corresponded to what the caregiver was experiencing at the time:

\begin{quote}
\small
``If I said that I felt primarily guilty today, then something that came up from the times when I felt proud might be useful.'' (P1)
\end{quote}

Together, these responses extend DG3 by showing how participants evaluated timely and personalized support through its fit with their immediate caregiving context. AI chatbot provided an accessible source of guidance, with participants distinguishing between general responses and interactions that felt more personally relevant. 
Likewise, the perceived usefulness of \cmr{}'s resources was shaped by their connection to caregivers' emotional experiences. Participants' responses therefore highlight contextual relevance as an important dimension of the timely and personalized support envisioned in DG3.


\subsubsection{Clinical Visibility Was Valuable When Paired with Meaningful Follow-Up}
\label{sec:6.1.5}


Participants generally responded positively to \cmr{}'s connection between the caregiver- and clinician-facing interfaces. 
In particular, making caregiver wellbeing visible through the clinician portal was seen as potentially useful when it could give clinician greater context about caregivers' experiences and help bring needs that might otherwise remain outside of clinical attention into view. P1 explained that the notes accompanying their mood reports could help clinicians interpret what broad wellbeing labels meant for them.
\begin{quote}
\small
``It's useful for both me and the clinician to see it, because the notes provide more context. Labels like `great,' `good,' `okay,' `low,' or `struggling' can mean different things to different people.'' (P1)
\end{quote}


That said, participants' responses also revealed a tension in this design: the same connection that could increase clinician awareness could affect how comfortable caregivers felt documenting their experiences in \cmr{}.
This tension was particularly apparent in participants' responses to the possibility that information entered through the caregiver-facing interface could become visible to their clinician. 
P13 explained that awareness of this connection could influence how openly they interacted with \cmr{}: \textit{``If it was just automatically reporting things back to my doctor, I might be less willing to have an honest conversation with it.''} 
P13's response highlights a potential trade-off within DG4. While connecting the two interfaces could make caregiver wellbeing more visible within clinical care, this visibility could also change the caregiver-facing space itself by making some caregivers more cautious about what they disclosed. 


Participants also evaluated clinical visibility in relation to what happened after caregiver needs became visible. Rather than viewing the clinician portal simply as a place for displaying caregiver information, participants saw greater value in this connection when clinicians could use the information to recognize when follow-up might be appropriate. P3 similarly saw value in clinician follow-up when caregivers needed it, while emphasizing that such attention should remain sensitive to the caregiver's immediate needs: \textit{``Sometimes it might be okay for the clinician to check up, but there are times we just want to be left alone.''} (P3)

Together, these responses both supported and complicated DG4. 
Participants saw value in making caregiver wellbeing more visible to clinicians, particularly when that visibility could support meaningful follow-up. 
At the same time, connecting caregiver reflections with clinical care introduced a tension between clinician awareness and caregivers' willingness to disclose sensitive experiences. 
These responses suggest that the value of \cmr{}'s clinician-facing features depends not simply on making more caregiver information visible, but on whether that visibility can support clinical attention without undermining the caregiver-facing space for reflection.

\subsection{RQ2: Envisioning a Clinically-connected Caregiver Wellbeing Ecosystem}
\label{sec:caregiver-envision}
Beyond participants' responses to existing \cmr{} features, participants reflected on how caregiver wellbeing support could be connected with clinical care more broadly. 
These responses revealed expectations and boundaries around the roles of caregivers, clinicians, and technology within a clinically connected wellbeing ecosystem. We organize these findings around four themes that capture how participants envisioned such a connection. 

\subsubsection{Caregiver Wellbeing Belongs within Dementia Care, but Should Remain Caregiver-Centered}
\label{sec:6.2.1}
Participants reflected more broadly on why caregiver wellbeing should receive attention within dementia care. Participants recognized that the wellbeing of caregivers and care recipients was closely interconnected. 
P10 emphasized the importance of attending to caregivers alongside the person living with dementia:

\begin{quote}
\small
``It's important to think about the patients, but if the caregiving is not [well supported] ... it's the patient who suffers, ultimately.'' (P10)
\end{quote}

This framing reveals an important tension in bringing caregiver wellbeing into the dementia care ecosystem. The effects of caregiver wellbeing on the care recipient provided one reason for clinical systems to pay greater attention to caregivers. At the same time, participants emphasized that caregiver wellbeing should also be supported for the caregiver's own sake. P4 explicitly distinguished these two beneficiaries of caregiver support:

\begin{quote}
\small
``I see this as an opportunity not just for mental health management and doing caregiving better, but for everyone---for the receiver and for the caregiver themselves.'' (P4)
\end{quote}

Therefore, participants' accounts suggest that connecting caregiver wellbeing with dementia care should \textit{not} position caregiver support solely as a means of improving care for the care recipient. 
While caregivers' and care recipients' wellbeing are intertwined, caregivers also have emotional and support needs that warrant attention independently of their effects on caregiving. 
\textit{A clinically connected caregiver wellbeing ecosystem should recognize caregivers both as partners within dementia care and as individuals whose wellbeing is itself an object of care.
}
\subsubsection{Making Caregiver Wellbeing Visible Creates a Responsibility to Respond}
\label{sec:6.2.2}
Participants first emphasized that making wellbeing visible required more than simply displaying a mood state; the representation needed to clearly communicate what that state meant. 
P14, for example, found the current color-coded feedback insufficiently distinct for conveying differences in wellbeing:

\begin{quote}
\small
``I would expect `great' to be green, and `struggling' to be really red. Right now, it's hard to tell whether those colors actually help communicate how you're doing.'' (P14)
\end{quote}

For participants, clearer representation could also help situate an individual mood within a broader pattern. P5, for example, envisioned comparing wellbeing across days: \textit{``It'd be neat to see, like, a chart, to see how it compares to another day.''} 
Making these changes easier to recognize also raised questions about what should happen when the system surfaced a concerning pattern. 
Participants envisioned future mechanisms through which sustained or substantial distress could lead to an appropriate next step, rather than remaining only as information displayed within \cmr{}. 
In particular, they described opportunities for the system to connect concerning wellbeing patterns with additional support or clinical attention.

\begin{quote}
\small

``If you've shown signs of significant distress, [the system could] show a chart that verifies it, and then say, `There are resources you can reach out to.''' (P1)
\end{quote}

These suggestions indicate that participants did not view greater visibility of caregiver wellbeing as an endpoint. Instead, they envisioned a progression from making wellbeing interpretable, to recognizing changes over time, to responding when those changes suggested a need for additional support. 
This expectation introduces a broader responsibility for a clinically connected caregiver wellbeing ecosystem: \textit{once caregiver distress becomes visible, the system must also consider when a response is warranted, who should respond, and what form of support should follow.}

\subsubsection{Connecting Caregivers and Clinicians Requires Negotiated Data Boundaries}
\label{sec:6.2.3}
Participants emphasized that connecting their wellbeing to clinical care, should not imply unrestricted or automated data sharing. 
Participants were generally open to clinicians receiving information that could help them understand caregiver wellbeing, while wanting to remain involved in deciding when personal information moved from the caregiver-facing space into clinical care. 
In particular, several participants described a permission-based model in which \cmr{} could prepare information for sharing while leaving the final decision to the caregiver.
For example, P13 described wanting \cmr{} to ask for permission before sharing information with a clinician:

\begin{quote}
\small
``Do you want me to share this with your doctor or something? Like, giving someone the choice might make me feel more comfortable about it.'' (P13)
\end{quote}

Such a permission step was important because participants viewed the caregiver-facing platform as a space, including the chatbot, where they might disclose experiences differently from how they would communicate with a clinician. 
P13 explained that automatic sharing could make them more cautious about what they disclosed to \cmr{}'s chatbot:

\begin{quote}
\small
``If it was just automatically reporting things back to my doctor, I might be less willing to have an honest conversation with it.'' (P13)
\end{quote}

Participants also envisioned these boundaries as extending beyond a single decision to share or not share. P8 described establishing criteria around what types of information could move to clinicians, suggesting that these preferences could be specified when caregivers first began using the system:

\begin{quote}
\small
``There should be something in the list, like, DO's and DON'Ts [..] the criteria which can be shared to the clinician, maybe in the initial setup.'' (P8)
\end{quote}

Together, these responses suggest that caregivers did not view clinical connection as a binary choice between sharing and keeping their wellbeing information private. 
Instead, they envisioned data sharing as an ongoing negotiation in which they could establish boundaries around what information crossed from \cmr{} into clinical care and retain agency at the moment of sharing. 
\textit{A permission-based sharing mechanism could make these boundaries visible within the interaction itself, allowing \cmr{} to facilitate caregiver--clinician communication without making clinical access the default consequence of personal reflection.}

\subsubsection{Technology Should Support Rather than Mediate the Caregiver--Clinician Relationship}
\label{sec:6.2.4}
Participants also considered the role that AI should play once caregiver wellbeing information became connected with clinical care. 
While AI-generated summaries could reduce the effort required to communicate caregivers' experiences, participants were concerned about relying on automated interpretations when those interpretations could influence how clinicians understood them. 
P13 drew on their awareness of AI-assisted clinical documentation to describe the potential consequences of such misinterpretation:

\begin{quote}
\small
``I've heard about some doctors using AI note-takers, and there have been cases where they've misinterpreted something or misheard something, and then a different diagnosis ends up in the patient's records because it misunderstood something.'' (P13)
\end{quote}

This concern became particularly salient when participants considered AI-generated summaries becoming part of clinical communication. P13 specifically worried about inaccurate or hallucinated information entering clinical records:

\begin{quote}
\small
``Something incorrect or [a] hallucination ... [could make] its way back into the records.'' (P13)
\end{quote}

Rather than relying on AI alone, P13 expressed greater trust in technology when a clinical professional remained involved in the process and could exercise professional judgment.
This preference for human involvement extended beyond concerns about accuracy. 
Participants also emphasized that caregiver wellbeing often involved deeply personal experiences for which technological interaction could not fully substitute for human support. As P14 explained:

\begin{quote}
\small
``[I would want an] answer from a person, not from a technical device, some robot type of thing. Because [..] those are really personal issues that people are trying to deal with.'' (P14)
\end{quote}

Together, these responses suggest that participants envisioned AI as supporting, rather than mediating, the caregiver--clinician relationship. 
AI could help summarize caregiver experiences and make wellbeing information easier to communicate, while participants expected human judgment to remain central to interpreting that information and determining how to respond. 
Their concerns about misinterpretation and hallucination further suggest that clinically connected AI should not become an autonomous voice speaking on behalf of caregivers. 
\textit{Therefore, it is critical to preserve a direct relationship between caregivers and clinicians, with AI serving as a tool that supports communication rather than replacing either caregiver expression or professional judgment.}

\section{Discussion}\label{sec:discussion}
In this work, we built and used \cmr{} as a design probe to examine how dementia caregivers perceive technologies designed to support their own wellbeing and how they envision such support connecting with the broader dementia care ecosystem. 
Across interviews with 14 caregivers, participants valued having an explicit space for caregiver wellbeing, following changes in their wellbeing over time, and receiving support that reflected their current caregiving circumstances. 
Our findings show that caregivers valued having a dedicated space for their own wellbeing, longitudinal awareness of how that wellbeing changed over time, and support that could reflect their current caregiving circumstances. 
At the same time, participants identified tensions around the burden and emotional consequences of continued reflection, the implications of making caregiver wellbeing visible within clinical care, the boundaries around what information should be shared, and the role of AI in interpreting and communicating caregivers' experiences. 
\autoref{tab:dg-ecosystem-mapping} synthesizes how these responses extended the four design goals of \cmr{}, showing how caregiver-facing wellbeing support shifts when it becomes part of a broader caregiver--clinician ecosystem. 
Across these shifts, our findings suggest that bringing caregiver wellbeing into dementia care is not only a question of providing additional support, but also of reconsidering how caregiver wellbeing is recognized, interpreted, and connected with clinical care. 
We discuss the implications of this work below.

\begin{table*}[t]
  \centering
    \footnotesize
    \sffamily
    \setlength{\tabcolsep}{3pt}
        \caption{Mapping the design goals of \cmr{} across caregiver-facing wellbeing support (\autoref{sec:caregiver-response}) and caregiver--clinician connection (\autoref{sec:caregiver-envision}). The mapping illustrates how each design goal evolves when personal wellbeing technologies become part of a broader caregiver--clinician ecosystem.}
    \label{tab:dg-ecosystem-mapping}
\begin{tabular}{p{0.19\columnwidth}p{0.24\columnwidth}p{0.25\columnwidth}p{0.24\columnwidth}}    
    \textbf{Design Goal (DG)}
    & \textbf{Caregiver-Facing Design}
    & \textbf{Caregiver--Clinician Connection}
    & \textbf{Conceptual Shift} \\
    \toprule

    \textbf{DG1:} \textbf{Open a Caregiver Wellbeing Pathway}
    \autoref{sec:dg-clinical-entry}
    &
    Create an explicit space for caregivers to attend to and reflect on their own wellbeing (\autoref{sec:6.1.2})
    &
    Make caregiver wellbeing visible as a legitimate concern within dementia care, rather than only through its effects on the care recipient (\autoref{sec:6.2.1}).
    &
    From recognizing caregiver wellbeing individually to positioning it as part of the broader care ecosystem.
    \\

    \hdashline

    \textbf{DG2: }\textbf{Translate Wellbeing Data into Interpretable Insights}
    \autoref{sec:dg-longitudinal}
    &
    Support longitudinal reflection through trends and AI-synthesized insights while accounting for the emotional consequences of revisiting past experiences (\autoref{sec:6.1.3}).
    &
    Surface meaningful changes and patterns that can support caregiver communication and clinician interpretation and response (\autoref{sec:6.2.2}).
    &
    From making wellbeing visible to connecting \emph{reflection}, \emph{interpretation}, and \emph{response}.
    \\

\hdashline
    \textbf{DG3: }\textbf{Provide Timely, Personalized, and Low-Burden Support}
    \autoref{sec:dg-personalized-support}
    &
    Provide personalized resources and support without turning wellbeing tracking or self-care into another caregiving task (\autoref{sec:6.1.4}).
    &
    Create pathways for additional support or follow-up when caregiver needs exceed what self-guided resources can address (\autoref{sec:6.2.2}).
    &
    From providing more support to reducing the work caregivers must perform to access appropriate support.
    \\

\hdashline
   \textbf{DG4:} \textbf{Preserve Privacy, Agency, and Data Boundaries}
    \autoref{sec:dg-clinician-followup}
    &
    Preserve a private space for reflection and give caregivers control over sensitive wellbeing information (\autoref{sec:6.1.5}).
    &
    Make clinical sharing permission-based and configurable, with caregivers determining what is shared, with whom, and when (\autoref{sec:6.2.3}).
    &
    From data privacy as protection to caregiver agency as an ongoing negotiation of clinical visibility.
    \\

    \bottomrule
    \Description{Mapping of \cmr{}'s four design goals across caregiver-facing wellbeing support and caregiver--clinician connection. DG1 extends an individual space for caregiver wellbeing into recognition of caregiver wellbeing within the broader care ecosystem. DG2 connects longitudinal reflection with interpretation and response. DG3 connects personalized, low-burden support with pathways to additional support or clinical follow-up. DG4 extends privacy and control toward permission-based and configurable clinical sharing.}
    \end{tabular}

\end{table*}

\subsection{Repositioning Caregiver Wellbeing within Dementia Care}

\subsubsection{From Care Partners to Recipients of Care}


Family caregivers are often positioned as essential in dementia care, and prior work has documented how caregiver wellbeing can affect both their capacity to provide care and outcomes for people living with dementia~\cite{schulz2004family, kim2021exploring}. 
Our findings extend this framing by showing that caregiver wellbeing should be valued not only because it could affect care for the person living with dementia, but because caregivers themselves warranted recognition and support.
This distinction shifts caregivers from being understood primarily as care partners toward also being recognized as a care recipient. 
For caregiver wellbeing technologies, this means designing beyond tools that help caregivers sustain their caregiving role or improve outcomes for the person receiving care. 

\subsubsection{From Tracking to Care: Making Wellbeing Visible Creates a Responsibility to Respond}


Our findings highlight the emotional consequences of making caregiver wellbeing visible over time. 
Longitudinal records allow caregivers to recognize patterns and contextualize changes in their wellbeing, while also resurfacing experiences that may remain emotionally difficult (\autoref{sec:6.1.3}). 
This makes wellbeing histories qualitatively different from neutral records of past behavior: what a system preserves and resurfaces can shape how caregivers revisit and make sense of difficult periods in their lives. 
While prior work on personal informatics has emphasized the value of longitudinal data for supporting self-knowledge and reflection~\cite{li2010stage,epstein2015lived}, our findings foreground the emotional weight that such reflection can carry in the context of caregiving. 

Making caregiver wellbeing visible also creates expectations for what follows. Participants envisioned concerning patterns as opportunities for interpretation, resources, and clinical follow-up (\autoref{sec:6.2.2}). This extends beyond tracking toward a progression from \emph{visibility}, to \emph{interpretation}, to \emph{response}: longitudinal data make changes recognizable, interpretation helps caregivers and clinicians understand their significance, and response connects those insights with appropriate forms of support. This progression resonates with broader work in personal health informatics emphasizing the importance of translating collected health data into information that can support reflection and action~\cite{li2010stage,epstein2015lived, murnane2018personal}. For caregiver wellbeing ecosystems, the value of longitudinal data therefore lies not simply in documenting how caregivers are doing, but in creating pathways through which emerging needs can be understood and acted upon.

\subsection{Negotiating Caregiver--Clinician Connection}

\subsubsection{Clinical Connection Should Be Negotiated, Not Assumed}

Connecting caregiver wellbeing technologies with clinical care changes the meaning of the information caregivers provide. Our findings reveal a tension between the benefits of clinical visibility and caregivers' willingness to disclose candidly: information that supports private reflection may feel different once caregivers anticipate that it could be seen by a clinician (\autoref{sec:6.2.3}). This distinction is particularly consequential for wellbeing technologies, where reflections may capture emotionally sensitive, uncertain, or momentary experiences. 
This concern parallels prior work on the observer effect, which shows that awareness of being monitored can itself alter how people behave and self-present in digital settings~\cite{saha2024observer,mccambridge2014systematic}.
Prior work on patient-generated health data has similarly emphasized the importance of understanding what people are willing to share and how such information moves across personal and clinical contexts~\cite{tiase2020patient,turner2021sharing}. Uncertainty about the consequences of disclosure may also shape what caregivers choose to report. If caregivers do not know what clinicians can see, what information may be flagged for attention, or what follow-up may occur, they may withhold or minimize experiences they would otherwise report. 

Our findings point toward clinical connection as a process of negotiated disclosure. Participants envisioned permission-based sharing and different boundaries for different forms of wellbeing information (\autoref{sec:6.2.3}), suggesting possibilities for graduated disclosure in which caregivers can move selected trends, summaries, or concerns into clinical care without exposing the entirety of their reflective space. 
Sharing preferences may evolve as caregiving circumstances, relationships with clinicians, and support needs change.

\subsubsection{AI Should Support, Rather than Mediate, the Care Relationship}

Our findings position AI as most useful when its role is grounded in the context of caregivers' existing needs and relationships. Participants valued AI-generated summaries for synthesizing longitudinal information and helping them recognize patterns that could otherwise be difficult to articulate (\autoref{sec:6.1.3}). However, within a caregiver--clinician ecosystem, these summaries become representations of caregivers' experiences that may shape how those experiences are understood by others (\autoref{sec:6.2.4}). Prior HCI work has similarly raised questions about how AI-mediated communication can shape communication and the interpretation of personal experiences~\cite{hancock2020ai}. AI-generated summaries therefore carry significance beyond their accuracy as summaries: they participate in representing the caregiver within a care relationship. 
This raises particular concerns when AI-generated interpretations omit context, misrepresent an experience, or introduce information through hallucination~\cite{pak2025polite}. 
In this framing, AI can reduce the work of synthesizing and communicating longitudinal wellbeing information while preserving the caregiver's voice and the clinician's judgment as central to the care relationship.

\subsubsection{Clinical Visibility Does Not Create Clinical Capacity}

Bringing caregiver wellbeing into the dementia care ecosystem also raises a practical tension between making needs visible and having the capacity to respond to them.
\cmr{} envisions pathways through which caregiver wellbeing information could inform clinical attention and follow-up, but greater visibility does not itself create additional clinical capacity.
Clinicians, social workers, and other support professionals already operate under substantial time and resource constraints, and introducing another stream of wellbeing information may create new responsibilities without providing the resources needed to fulfill them.
A clinically connected caregiver wellbeing ecosystem therefore cannot assume that identifying additional caregiver needs will automatically translate into timely human support.

This also raises unresolved questions about accountability.
If a system identifies concerning distress, who is responsible for reviewing that information, determining whether a response is warranted, and ensuring that follow-up occurs?
These questions become particularly consequential if caregiver wellbeing information is incorporated into clinical workflows, where ambiguity around responsibility may carry both organizational and potential liability implications.
Our study does not resolve these questions, and answering them will require engagement with clinicians, health systems, social workers, and other stakeholders who would be responsible for operationalizing such a model of care.

Resource constraints also raise questions about equitable access.
For example, if clinically connected caregiver support were offered as a pay-per-use service, caregivers with fewer financial resources would be disadvantaged.
Conversely, limiting such support to caregivers who show substantial distress could conserve scarce clinical resources, but would move the system away from its proactive model.
Therefore, determining how caregiver wellbeing support should be funded, prioritized, and integrated into existing care infrastructures will require future deployment and efficacy studies alongside consideration of clinical capacity, costs, reimbursement models, and health policy.

\subsection{Design Implications for Caregiver Wellbeing Ecosystems}

\subsubsection{Design wellbeing support that does not become another caregiving responsibility.}
Our findings showed that support intended for caregivers can itself become another demand on their limited time and attention. 
As discussed in \autoref{sec:6.1.2}, participants valued having an explicit space for their own wellbeing, yet repeated assessments and check-ins could also feel like another obligation. 
These findings suggest that future caregiver wellbeing systems should consider not only what support they provide, but also the additional work caregivers must perform to receive it. 
Systems can reduce this burden through brief and skippable interactions, adaptable check-in frequency, and support surfaced in response to caregivers' current needs. 
More broadly, proactive caregiver wellbeing technologies can reduce the work of seeking support rather than turn self-care into another responsibility caregivers are expected to manage~\cite{epstein2015lived,li2010stage}.

\subsubsection{Treat longitudinal wellbeing data as emotionally consequential and couple visibility with pathways to response.}
\autoref{sec:6.1.3} notes how participants valued being able to revisit previous experiences and contextualize their current wellbeing, while also describing how resurfacing difficult moments could reinforce distress.
Longitudinal wellbeing data should therefore not be treated simply as a neutral history to be collected and repeatedly surfaced.
Participants further expected greater visibility of wellbeing to lead toward meaningful support when concerning patterns emerged (\autoref{sec:6.2.2}). 
Therefore, making caregiver wellbeing visible should not itself be the endpoint of longitudinal tracking. Future caregiver wellbeing systems should consider both \emph{when} wellbeing histories are surfaced and \emph{what happens next}. Systems can provide caregivers with greater control over how and when past experiences are revisited, while connecting sustained or concerning changes with appropriate resources or opportunities for clinical follow-up. 
In this way, longitudinal wellbeing technologies can move beyond simply making wellbeing visible toward supporting a progression from \emph{reflection}, to \emph{interpretation}, to \emph{response}.

\subsubsection{Preserve private reflection while making clinical sharing permission-based and configurable.}
Our findings revealed that privacy is not only a question of protecting caregiver data, but also of preserving a space in which caregivers can reflect openly without anticipating clinical visibility.
As discussed in \autoref{sec:6.1.5}, participants valued \cmr{} as a space for recognizing and reflecting on their own wellbeing. 
Yet when this information became connected to clinical care, participants drew clearer boundaries around what should remain private and what could be shared (\autoref{sec:6.2.3}).
Caregiver wellbeing ecosystems should preserve a meaningful distinction between information used for private reflection and information made available to clinical care.
Systems could provide caregivers with mechanisms to share selected summaries, trends, or concerns without making the entirety of their reflective space clinically visible.
Caregivers should be able to determine what information is shared, with whom, and under what circumstances, and to revise these preferences as their needs and caregiving situations change.
Such mechanisms allow caregiver wellbeing to inform clinical care without treating participation in wellbeing support as permission for continuous clinical access.

\subsubsection{Use AI to support caregiver expression and clinician judgment rather than mediate the relationship.}
A key design implication concerns how future caregiver wellbeing ecosystems should position AI as infrastructure for communication and sense-making rather than as a substitute for caregiver--clinician interaction.
Caregivers should be able to inspect, contextualize, and correct AI-generated representations before they are shared, while clinicians should be able to distinguish caregiver-reported information from AI-generated summaries or inferences.
AI can thereby reduce the work of synthesizing and communicating longitudinal wellbeing information without transferring interpretive authority away from caregivers or replacing clinician judgment.
At the same time, AI-generated interpretations should remain visible to caregivers and subject to clinician judgment rather than being treated as definitive assessments or automated decisions. 
Such a role allows AI to reduce the work of making caregiver experiences legible within clinical care while keeping caregiver expression and clinician judgment at the center of the relationship.

\subsection{Limitations and Future Directions}
Our work has limitations, which also suggest future directions. First, the caregiver participants in our study are not a representative sample of the broader dementia caregiver population, particularly given our relatively small sample of recruitment through online communities. However, our primary goal is not to produce generalizable findings, but rather to understand caregivers' expectations and design preferences and use these insights to inform the design of an ecosystem connecting caregivers and clinicians around caregiver wellbeing. Then, our work did not include clinicians as study participants. Thus, while our findings capture caregivers' perspectives on sharing wellbeing information and connecting \cmr{} with clinical care, they do not reflect clinicians' needs, concerns, or existing clinical workflow. This is particularly important because the envisioned ecosystem depends on clinicians interpreting caregiver wellbeing information and determining when and how to response. Participants' responses therefore reflect anticipated rather than sustained real-world use.
Future work can extend \cmr{} by engaging both caregivers and clinicians to examine how clinicians interpret wellbeing scores and indicators of potential concern, what they consider actionable, and what forms of follow-up are feasible within existing clinical roles and workflows. 
Longitudinal deployments could further examine how caregivers’ needs and preferences evolve over time, including their engagement with mood tracking, AI-synthesized insights, and personalized support, as well as whether these features remain useful without introducing additional burden. Future studies could also recruit caregivers from more diverse backgrounds and care contexts and investigate how factors such as caregiving stage, access to healthcare, and digital literacy shape expectations for proactive and clinically connected wellbeing support.

\section{Conclusion}
In this work, we presented \cmr{}, an envisioned proactive wellbeing ecosystem that explores how caregiver wellbeing support might be integrated into the broader dementia care ecosystem. Through semi-structured interviews with 14 caregivers, we found that caregivers valued technologies that could help them recognize changes in their wellbeing over time, make sense of those changes, and access timely and personalized support. At the same time, connecting caregiver wellbeing with clinical care introduced important tensions around privacy, disclosure, emotional burden, and caregiver agency. Our findings suggest that making caregiver wellbeing visible should not be an endpoint: visibility should be accompanied by meaningful interpretation and pathways to response, while caregivers retain control over what enters clinical contexts and when. More broadly, this work highlights an opportunity to design caregiver--clinician ecosystem that recognize caregivers not only in relation to the people they support, but as individuals with wellbeing needs of their own. 


\begin{acks}
We sincerely thank all study participants for their time, insights, and willingness to share their experiences. This work was partly supported by the Jump ARCHES endowment through the Health Care Engineering Systems Center at Illinois and the OSF Foundation.
This work was also partly supported by the National Institute on Aging of the National Institutes of Health under Award Number P30AG073105. 
\end{acks}

\bibliographystyle{ACM-Reference-Format}
\bibliography{references}



\end{document}

\endinput